\documentclass[lineno]{JFM-FLM_Au}

\usepackage{upgreek}
\newcommand\ohn{\mbox{\textit{Oh}}}
\newcommand{\iu}{\mathrm{i}}
\newcommand{\eu}{\mathrm{e}}
\newcommand{\du}{\mathrm{d}}

\lefttitle{Y. Mao, C. Zhao, Y. Zhang, K. Mu and T. Si}
\righttitle{Influence of vdW forces on film instability in a tube}

\title{Influence of van der Waals forces on the instability of a liquid film in a tube}

\author{Yixiao Mao\aff{1}, Chengxi Zhao\aff{1}, Yixin Zhang\aff{2}, Kai Mu\aff{1} \and Ting Si\aff{1}}

\affiliation{\aff{1}Department of Modern Mechanics, University of Science and Technology of China, Hefei 230026, PR China \aff{2}Physics of Fluids Group and Max Planck Center Twente for Complex Fluid Dynamics, MESA+ Institute and J. M. Burgers Centre for Fluid Dynamics, University of Twente, P.O. Box 217, 7500 AE Enschede, The Netherlands}

\corresau{Chengxi Zhao, \email{zhaochengxi@ustc.edu.cn}; Ting Si, \email{tsi@ustc.edu.cn}}

\begin{document}
\maketitle

\begin{abstract}
The instability of a liquid film in a nanotube is significantly influenced by van der Waals forces. A theoretical framework based on the axisymmetric Stokes equations is developed to investigate their effects through linear stability analysis. The model reveals that van der Waals forces markedly enhance perturbation growth, reduce the dominant wavelength, and lower the critical film thickness that distinguishes collapse from non-collapse regimes. Direct numerical simulations of the Navier-Stokes equations both confirm these theoretical predictions and extend the analysis into the nonlinear regime. In this regime, van der Waals forces are found to alter the interfacial morphology and suppress the formation of satellite lobes. Both rupture and collapse follow a universal temporal scaling law with exponent 1/3 and exhibit self-similar behavior near the singularity. 

\end{abstract}



\section{Introduction}
\label{sec:headings}

Thin liquid films in capillaries are ubiquitous in both natural systems and industrial processes. 
Examples include fluid transport in xylem conduits \citep{jacobsen2024} and multiphase flow in porous geological formations \citep{Ben‐noah2023}.
Substantial research has also been devoted to pulmonary airway closure—a pathological process driven by instability in the mucus layer, which is associated with respiratory conditions such as asthma and respiratory distress syndrome \citep{Heil2008,bian2010,Erken2022}.
In engineering contexts, film stability critically influences applications such as fuel cell water management \citep{lu2011}, enhanced oil recovery via flooding techniques \citep{olbricht1996,Perazzo2018}, and controlled interfacial patterning in microfluidic devices \citep{zoueshtiagh2014}.
Given these diverse contexts, accurate prediction of thin film instability evolution carries significant scientific and practical importance.

Within the classical theoretical framework, an axisymmetric liquid film coating the interior of a capillary tube, governed primarily by surface tension, tends to minimize its surface energy through deformation \citep{Gennes2004}. The dominant instability mechanism arises from the Rayleigh--Plateau instability \citep{plateau1873,pri}, which reflects a competition between the two principal curvatures: the radial component promotes destabilization, while the axial component provides a  stabilizing influence. As a result, the interface becomes unstable to long-wavelength perturbations exceeding a critical threshold.
\cite{pri} conducted a foundational study on the breakup of liquid jets into droplets, and established the corresponding linear stability theory. Linear stability analysis indicates a cutoff wavelength of $\lambda_c = 2\pi r_f$, corresponding to a dimensionless critical wavenumber $k_{c} = 1$, where $r_f$ is the unperturbed radius of the liquid film interface. The dominant mode corresponding to the maximum growth rate occurs at $2\pi r_f / \lambda = 0.696$.
Subsequently, \cite{goren1961} extended the jet-breakup theory to describe the instability of liquid films coating the interior or exterior of cylindrical capillaries. This analysis confirmed that such films also exhibit periodic undulations, analogous to the breakup of free jets. Although the cutoff wavelength $\lambda_c$ remains identical to that of a liquid column, the presence of the wall modifies both the dominant wavelength and the growth rate of perturbations in the linear regime. Experimental observations of droplet formation on cylindrical fibers and within tubes have shown good agreement with these theoretical predictions.

The nonlinear evolution of liquid films leads to two distinct configurations, depending on the film volume within a single wavelength: wavy collars at small volumes and liquid plugs at large volumes. Through static thermodynamic analysis, \cite{everett1972} established the existence of a critical film volume, below which collapse is prevented due to volume constraints. \cite{hammond1983} developed a lubrication model for viscous-dominated flows to study nonlinear film deformation, capturing the long-term evolution of collars, and predicting the emergence of secondary maxima (satellite lobes) along with subsequent drainage processes. 
\cite{Gauglitz1988} extended Hammond's model by incorporating more precise surface tension boundary conditions, deriving a critical dimensionless film thickness $\epsilon_c =1- r_f / r_0 = 0.12$ that distinguishes non-collapsing collars from collapsing plugs, where $r_0$ is the tube radius. This threshold was subsequently validated experimentally. Further, \cite{Lister2006} investigated the long-term evolution of collars, revealing complex dynamics involving multiple collar interactions in long tubes, particularly for prolonged relaxation arising from coupling between large-amplitude collars and small-amplitude lobes.
Beyond the extensive development of lubrication-based models, incorporating additional physical factors introduces further complexity in both theoretical formulations and observed phenomena. Core fluid flow and axial oscillations have been shown to suppress channel closure \citep{frenkel1987,halpern2003}, while inertial effects tend to decelerate the instability development \citep{rykner2024}. When axial gravity acts as an external driving force, it may induce either absolute or convective instability, depending on specific flow conditions \citep{ruyer2008}.

With advances in nanoscale technology and growing application demands, liquid film dynamics at the nanoscale have garnered increasing research interest in areas such as crystal growth \citep{day2015} and fluid transport in carbon nanotubes \citep{falk2010}. Previous studies have shown that a range of small-scale physical effects, including slip boundaries \citep{liao2013} and thermal fluctuations \citep{Fetzer2007,zhang2021,zhao2022,zhang2024}, significantly influence the instability evolution of thin liquid films.
Intermolecular interactions, particularly van der Waals (vdW) forces, also play an important role in nanoscale film behavior. Their effects are strongly dependent on the material properties of the interacting interfaces \citep{israelachvili2011}.
In the case of completely wetting liquids and substrates, vdW forces typically manifest as attractive forces that promote the formation of stable liquid films. Examples include films deposited on fibers drawn from liquid baths \citep{quere1989} and stable layer flow between droplets on fibers \citep{ji2019,Calvo2025}.
Conversely, for non-wetting systems, disruptive intermolecular interactions can induce film rupture within finite time scales, which are well-established as a key mechanism driving instability and spinodal dewetting in liquid films on planar substrates \citep{Craster2009}.

Notably, recent experimental work has observed instability structures in ultra-thin water films confined within nanoscale single-layer graphene channels \citep{tomo2022}, where the dominant wavelength measured only 20--44\% of that predicted by classical instability theory. These observations were attributed to destabilizing vdW forces exerted by the solid walls on the liquid interface. A lubrication model based on the conventional thin-film assumption was employed to describe the underlying mechanism. The results demonstrate that the vdW term significantly reduces the fastest-growing wavelength and enhances the growth rate, particularly in liquid films thinner than $5\,\rm{nm}$.
The theoretical model further reveals that under vdW influence, the dominant wavelength $\lambda_{{max}}$ increases from zero to a maximum as film thickness grows, then decreases towards zero. This result raises questions regarding the reliability of lubrication models for thicker films. Indeed, such models remain valid only for extremely thin films in the linear regime \citep{zhao2023}, highlighting the need for a Stokes-based linear stability framework to accurately explore a broader parameter range.
Beyond linear instability, experiments by \cite{tomo2022} also captured nonlinear evolution, though detailed analysis was not provided. Previous studies indicate that vdW forces substantially affect nonlinear rupture dynamics for liquid films on a planar substrate, for example, the lubrication theory gives a scaling relation $h_{{min}} \propto \tau^{1/5}$ near the rupture point \citep{zhang1999}, where $h_{{min}}$ is the minimum film thickness and $\tau$ is the time to rupture. Recent studies, however, suggest that the flow follows a universal self-similar solution of the Stokes equations, yielding the scaling $h_{{min}} \propto \tau^{1/3}$ \citep{Moreno-Boza2020}.
It is also noteworthy that while \cite{tomo2022} focused on vdW forces between liquid and solid walls, intermolecular attraction between liquid--gas interfaces themselves plays a significant role. As liquid films approach closely, vdW attraction may trigger instantaneous contact \citep{Chireux2018,Beaty2022,Beaty2023}, yet how this mechanism influences film collapse remains a question requiring further investigation.

To address the issues above and capture interface evolution more accurately, this paper presents a combined theoretical and numerical investigation of the evolution of liquid films in cylindrical tubes under the influence of vdW forces. 
The specific contents are organized as follows: In \S\,\ref{sec:model}, the theoretical framework and instability analysis based on both Stokes and lubrication models are given in detail. The computational methodology is described in \S\,\ref{sec:simulation}. The influence of vdW forces on the two evolutionary modes, including collapse and non-collapse, is examined in \S\,\ref{sec:two mode}. The dominant wavelength and growth rate during the linear evolution stage are discussed and validated in \S\,\ref{sec:linear}. The morphological differences and singularity scaling laws associated with rupture and collapse in the nonlinear stage are analyzed in \S\,\ref{sec:nonlinear}. Conclusions are summarized in \S\,\ref{sec:conclusion}.

\section{Mathematical Model}
\label{sec:model}
In this section, a mathematical model for the dynamics of a liquid film inside a cylindrical tube under the influence of vdW forces is developed. Specifically, the governing equations for this configuration are presented in  \S\,\ref{sec:Model formulation}, and a linear stability analysis is provided in \S\,\ref{sec:dispersion}, comparing the dispersion relations between the Stokes and lubrication models.

\subsection{Model formulation}
\label{sec:Model formulation}

\begin{figure}
  \centerline{\includegraphics[width=0.9\linewidth]{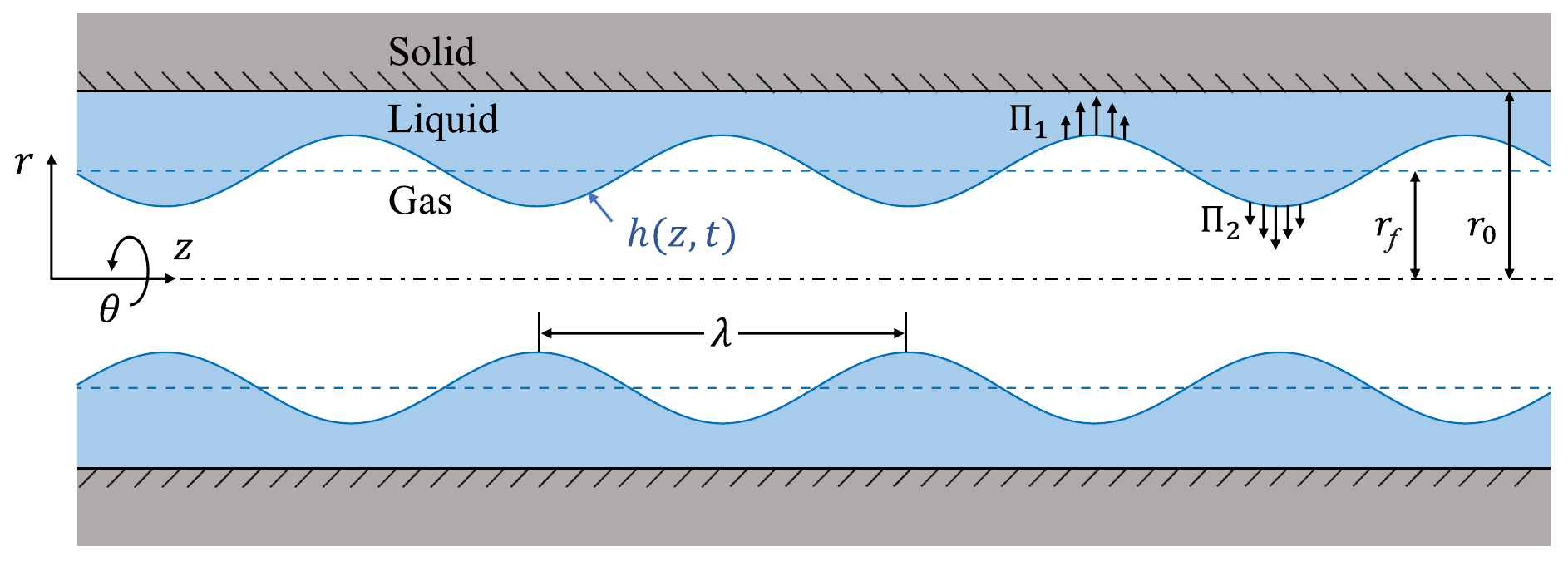}}
  \caption{Schematic of annular film flow in a tube.}
\label{fig:film}
\end{figure}

We consider a Newtonian liquid film coating the interior of a horizontal cylindrical tube of radius $r_0$, with the $z$-axis aligned along the tube centerline (see figure \ref{fig:film}). The initial radius of the liquid–gas interface from the $z$-axis is denoted as $r = r_f$. The flow dynamics within the liquid film is governed by the incompressible Navier–Stokes (NS) equations.
To simplify the governing parameters, we non-dimensionalise the NS equations using the following scaling:
\begin{eqnarray}
\label{eq_model1}
(r,z,h)=\frac{(\tilde{r},\tilde{z},\tilde{h})}{r_0},\quad
t=\frac{\gamma}{\mu r_0}\tilde{t},\quad
{\boldsymbol{u}}=\frac{\mu}{\gamma}\tilde{\boldsymbol{u}},\quad\quad\quad\nonumber\\
(p,\boldsymbol{\sigma},\Pi_1,\Pi_2)=\frac{r_0}{\gamma}(\tilde{p},\tilde{\boldsymbol{\sigma}}, \tilde{\Pi}_1, \tilde{\Pi}_2),\quad
(A_1,A_2)=\frac{(\tilde A_1,\tilde A_2)}{6\pi\gamma r_0^2},
\end{eqnarray}
where $\tilde h$, $\tilde t$, $\tilde{\boldsymbol{u}}$ and $\tilde p$ represent the dimensional interface height, time, velocity and pressure, respectively. (Note that dimensional material parameters are written without tildes.) Here, $\tilde{\boldsymbol{\sigma}}$ denotes the shear stress, which is proportional to the strain rate in Newtonian fluids, $\mu$ is the dynamic viscosity of the liquid, and $\gamma$ is the surface tension of the liquid–gas interface. 

The disjoining pressure induced by vdW forces is modeled via the Derjaguin approximation $\Pi = A/d^3$ \citep{deryagin1955}, where $A$ represents the dimensionless Hamaker constant, and $d$ denotes the separation between interfaces. Within this framework, the disjoining pressure at a location $z$ is equated to that between two infinite, flat, and parallel surfaces spaced by $d(z,t)$. As shown in figure \ref{fig:film}, two contributions to the disjoining pressure are taken into account: ${\Pi}_1$ (between the solid–liquid and liquid–gas interfaces) and ${\Pi}_2$ (between opposing liquid–gas interfaces). These correspond to the Hamaker constants ${A}_1 = {A}_{solid-liquid}$ and ${A}_2 = {A}_{liquid-liquid}$, which represent the relative strengths of vdW forces. Their values may span a broad range depending on material properties and system size.
For most condensed phases, the dimensional Hamaker constant $\tilde{A}$ typically ranges from $10^{-21}$ to $10^{-19}\ \mathrm{J}$ \citep{israelachvili2011}.
For example, the silicone-oil drops used in the post-contact coalescence experiments of \cite{Paulsen2012} have $\tilde A_2=10^{-20}\,\rm{J}$, $\gamma=0.02\,\rm{N \,m^{-1}}$, and $r_0 = 0.001\,\rm{m}$ \citep{Beaty2022}, which gives $A_2= 2.65 \times 10^{-14}$. The water films inside a graphene nanoscroll in experiments of \cite{tomo2022} have $\tilde A_1=1.82\times10^{-20}\,\rm{J}$, $\gamma=0.07\,\rm{N \, m^{-1}}$, and $r_0 = 13.0\text{--}33.9\,\rm{nm}$, which gives $A_1= 1.20\text{--}8.18 \times 10^{-5}$.
For cases with low interfacial tension, the value will be even higher. Therefore, we use a parameter range of $10^{-9}\text{--}1$ for $A_1$ and $A_2$, mainly focusing on the liquid film within nanoscale tubes.
The dimensionless NS equations are given by
\begin{align}
\label{eq_model2}
\nabla \cdot {\boldsymbol{u}}&=0 \,,\\
\label{eq_model3}
\frac{\partial u}{\partial t}+\boldsymbol u \cdot \nabla {\boldsymbol u} &= \ohn^2 (\nabla \cdot {\boldsymbol \sigma} - \nabla p)\, , 
\end{align}
where the Ohnesorge number $\ohn=\mu/\sqrt{\rho \gamma r_0}$ , represents the ratio of viscous forces to inertial and surface-tension forces. Under the assumption $\ohn^2\gg 1$, \color{black}inertial terms become negligible compared to viscous terms. For axisymmetric initial perturbations, (\ref{eq_model2}) and (\ref{eq_model3}) simplify to the axisymmetric Stokes equations:
\begin{align}
\label{eq_model4}
\frac{\partial w}{\partial z}&+ \frac{1}{r} \frac{\partial(ur)}{\partial r} =0\,, \\
\label{eq_model5}
\frac{\partial p}{\partial z}=& \frac{\partial^2 w}{\partial z^2}+ \frac{1}{r}\frac{\partial}{\partial r}\biggl(r \frac{\partial w}{\partial r}\biggr)\, ,\\ 
\label{eq_model6}
\frac{\partial p}{\partial r}=& \frac{\partial^2 u}{\partial z^2}+ \frac{\partial}{\partial r}\biggl[\frac{1}{r} \frac{\partial (wr)}{\partial r}\biggr]\,,
\end{align}
where $u$ and $w$ represent the radial and axial velocities, respectively. Since the density of gas surrounding the film is much lower than the liquid density, the gas flow can be treated as dynamically passive. The liquid–gas interface height $h(z,t)$ satisfies the kinematic boundary condition
\begin{equation}
\label{eq_model7}
    \frac{\partial h}{\partial t}+ w \frac{\partial h}{\partial z} =u\,. 
\end{equation}
At the interface $r=h$, the normal stress balance yields
\begin{equation}
\label{eq_model8}
    p-\boldsymbol{n\cdot \sigma \cdot n}= \nabla \boldsymbol \cdot \boldsymbol n +\Pi_1-\Pi_2\, ,
\end{equation}
where $\boldsymbol n$ is the outward normal unit vector, and $\nabla \cdot \boldsymbol n$ represents the dimensionless Laplace pressure. The disjoining pressures are $\Pi_1=A_1/(1-r)^3$ and $\Pi_2=A_2/(2r)^3$ respectively. The tangential stress balance is
\begin{equation}
\label{eq_model9}
    \boldsymbol{ n \cdot \sigma \cdot t}=0\,,
\end{equation}
with $\boldsymbol t$ as the tangential unit vector. Expressing $\boldsymbol n$ and $\boldsymbol t$ in terms of the Cartesian unit vectors $\hat{\boldsymbol e}_z$ and $\hat{\boldsymbol e}_r$ yields:
\begin{equation}
\label{eq_model10}
    {\boldsymbol n}=\frac{\partial_z h}{\sqrt{1+(\partial_z h)^2}}\hat{\boldsymbol e}_z-\frac{1}{\sqrt{1+(\partial_z h)^2}}\hat{\boldsymbol e}_r \,,\,
    {\boldsymbol t}=\frac{1}{\sqrt{1+(\partial_z h)^2}}\hat{\boldsymbol e}_z+\frac{\partial_z h}{\sqrt{1+(\partial_z h)^2}}\hat{\boldsymbol e}_r \,.
\end{equation}
Substituting (\ref{eq_model8}) and (\ref{eq_model9}) gives the explicit forms:
\begin{align}
\label{eq_model11}
p-\frac{2}{1+(\partial_z h)^2} \biggl[ \frac{\partial u}{\partial r}- \frac{\partial h}{\partial z}\biggl(\frac{\partial w}{\partial r}+\frac{\partial u}{\partial z} \biggr)+ \biggl( \frac{\partial h}{\partial z}\biggr)^2\frac{\partial w}{\partial z}\biggr]\nonumber\\
=\frac{\partial_z^2 h}{[1+(\partial_z h)^2]^{3/2}}-\frac{1}{h\sqrt{1+(\partial_z h)^2}} +\frac{A_1}{(1-h)^3} -\frac{A_2}{(2h)^3}\,, \\
\label{eq_model12}
2 \frac{\partial h}{\partial z}\biggl( \frac{\partial u}{\partial r}- \frac{\partial w}{\partial z}\biggr) +\biggl[1-\biggl( \frac{\partial h}{\partial z}\biggr)^2\biggr]\biggl( \frac{\partial w}{\partial r}+ \frac{\partial u}{\partial z}\biggr)=0\,. 
\end{align}
 At the tube wall $r=1$, no-slip and no-penetration conditions apply:
\begin{align}
\label{eq_bc1}
u = 0\,, \,w = 0 \,.
\end{align}

\subsection{Linear stability analysis}
\label{sec:dispersion}

In this subsection, we perform a linear instability analysis of equations (\ref{eq_model4})–(\ref{eq_bc1}) using the normal mode method. This approach has been widely employed in various interfacial instability problems, including the breakup of liquid columns \citep{tomotika1935} and the instability of liquid films on cylindrical fibers \citep{goren1961,zhao2023}.
The dimensionless perturbed quantities are expressed as
\begin{equation}
\label{eq_smallap}
u(r,z,t) = \hat{u}(r) \eu^{\omega t+\iu kz}, 
w(r,z,t) = \hat{w}(r) \eu^{\omega t+\iu kz}  \textrm{ and } 
p(r,z,t) = 1 + \hat{p}(r) \eu^{\omega t+\iu kz}\,,
\end{equation}
where $\omega$ is the perturbation growth rate and $k$ is the wavenumber.
Similarly, the interface profile is linearised using $h(z,t) = \alpha + \hat{h}\eu^{\omega t+\iu kz}$, where $\alpha=r_f/r_0$ is the dimensionless film position.
The dispersion relation between $\omega$ and $k$ can be expressed explicitly as
\begin{align}
\label{eq_dispersion_stokes}
\omega &= -\frac{1}{2}\left( k^2- \frac{1}{\alpha^2} - \frac{3A_1}{ (1-\alpha)^4} - \frac{3A_2}{8 \alpha^4} \right) \nonumber\\
& \frac{K_0(k \alpha) \alpha \Delta_1 - K_1(k \alpha) \Delta_2 + I_0(k \alpha) \alpha \Delta_3 -I_1(k \alpha) \Delta_4}{k \alpha K_1(k \alpha) \Delta_1- \left[ kK_0(k \alpha)+K_1(k \alpha)/\alpha \right] \Delta_2 -k \alpha I_1(k \alpha) \Delta_3 + 
\left[ k I_0(k \alpha) -I_1(k \alpha)/\alpha \right] \Delta_4} \,,
\end{align} 
where
\begin{align}
\Delta_1& = 
\begin{array}{|ccc|}
 -kK_0(k)  & 2 I_0(k)+k I_1(k) & kI_0(k) \\ 
 K_1(k) &  I_0(k) & I_1(k) \\
 k K_1(k\alpha) & k \alpha I_0(k \alpha) + I_1(k \alpha) & k I_1(k\alpha) \\
\end{array}\,, \nonumber \\ \nonumber
\Delta_2 &= 
\begin{array}{|ccc|}
2K_0(k)-kK_1(k)   & 2 I_0(k)+k I_1(k) & kI_0(k) \\ 
 K_0(k) &  I_0(k) & I_1(k) \\
k \alpha K_0(k \alpha) - K_1(k\alpha) & k \alpha I_0(k \alpha) + I_1(k \alpha) & k I_1(k\alpha) \\
\end{array}\,, \\ \nonumber
\Delta_3 &= 
\begin{array}{|ccc|}
2K_0(k)-kK_1(k) & -kK_0(k)  & kI_0(k) \\ 
 K_0(k) & K_1(k) & I_1(k) \\
k \alpha K_0(k \alpha) - K_1(k\alpha) & k K_1(k\alpha) & k I_1(k\alpha) \\
\end{array}\,, \\ \nonumber
\Delta_4 &= 
\begin{array}{|ccc|}
2K_0(k)-kK_1(k) & -kK_0(k)  & 2 I_0(k)+k I_1(k) \\ 
 K_0(k) & K_1(k) &  I_0(k) \\
k \alpha K_0(k \alpha) - K_1(k\alpha) & k K_1(k\alpha) & k \alpha I_0(k \alpha) + I_1(k \alpha)  \\
\end{array}\,.  
\end{align}
Here, $I_n$ and $K_n$ are modified Bessel functions of the first and second kind respectively, with subscripts denoting their order. The detailed derivation of this formulation is provided in Appendix \ref{appA}.

To compare the differences between the Stokes model and the thin-film-based lubrication model in describing this problem, we also derived the governing equations and corresponding dispersion relations for the lubrication model in a cylindrical tube. The complete derivation is presented in Appendix \ref{appB}, which yields:

\begin{equation}
\label{eq_le_dis}
    \omega=k^2 \biggl( k^2-\frac{1}{\alpha^2}-\frac{3A_1}{(1-\alpha)^4}-\frac{3A_2}{8\alpha^4} \biggr)M\,,
\end{equation}
where $M=(-1+4\alpha^2-3\alpha^4+4\alpha^4\ln \alpha)/16$.

The dispersion relation of the lubrication model from \cite{tomo2022} is
\begin{equation}
\omega=-\frac{1}{3}(1-\alpha)^3 k^2 \biggl( k^2-\frac{1}{\alpha^2}-\frac{3A_1}{(1-\alpha)^4}-\frac{3A_2}{8\alpha^4} \biggr)\,.
    \label{eq_lub_tomo}
\end{equation}
It should be noted that the original model only accounted for vdW forces exerted by a thin wall on the liquid film. In the present study, however, we assume an infinitely thick wall, and incorporate additional vdW forces arising from the liquid–liquid interactions. The model has been modified accordingly to reflect these physical considerations.

\begin{figure}
  \centerline{\includegraphics[width=1\linewidth]{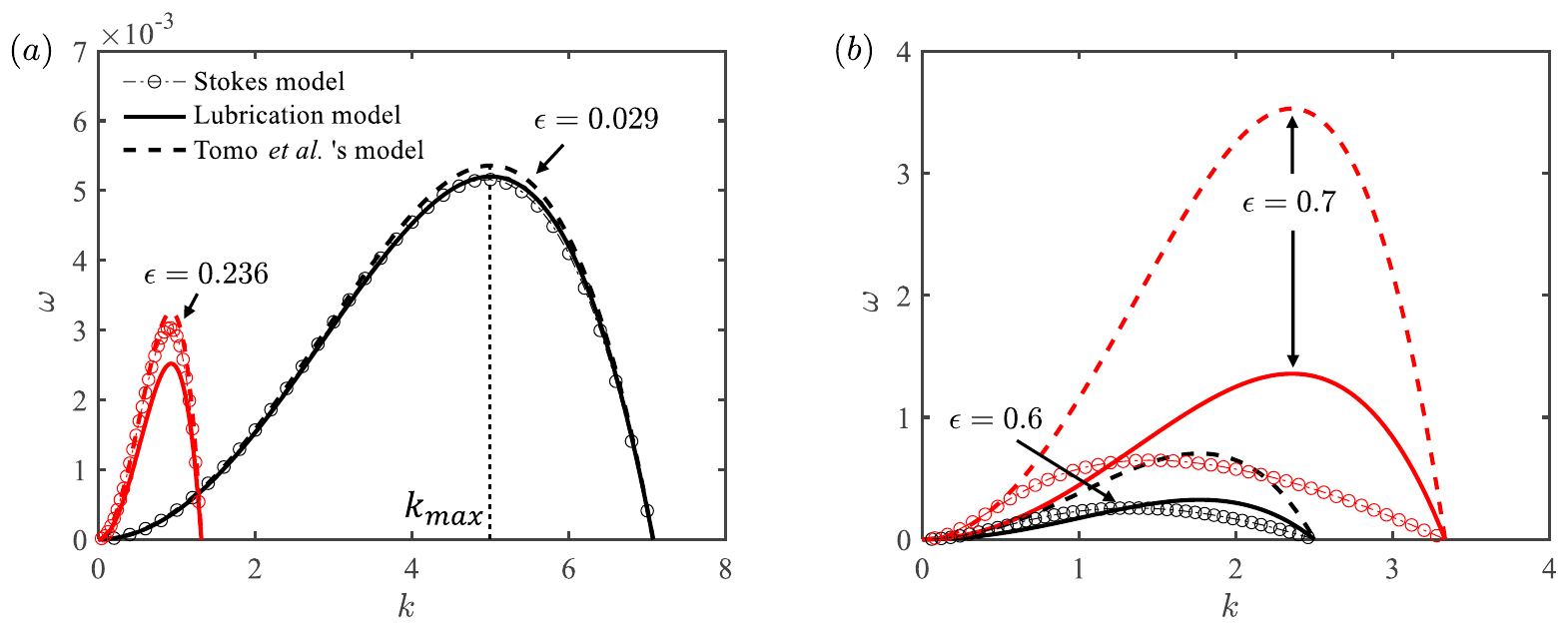}}
  \captionsetup{width=1\linewidth}
  \caption{The dispersion relation between the growth rate $\omega$ and the wavenumber $k$ with $A_1=1.2\times 10^{-5}$ of various film thickness: ($a$) $\epsilon=0.029,0.236$, ($b$) $\epsilon=0.6,0.7$. Circles with dash-dotted lines represent results of Stokes model (\ref{eq_dispersion_stokes}), solid lines represent results of lubrication model (\ref{eq_le_dis}), and dashed lines represent results of lubrication model (\ref{eq_lub_tomo}) by \cite{tomo2022}.}
\label{fig:wavenumber}
\end{figure}

Figure \ref{fig:wavenumber} compares the dispersion relations of the Stokes model, the lubrication model, and the model by \cite{tomo2022} under the influence of vdW forces.
To more intuitively represent the influence of the liquid film thickness, we define the dimensionless thickness $\epsilon = 1 - \alpha$ for subsequent investigation and analysis.
We adopt the same parameters as those in the experiments of \cite{tomo2022}, where $\tilde{A}_1 = 1.82 \times 10^{-20}\,\mathrm{J}$, $\gamma = 0.07\,\mathrm{N\,m^{-1}}$, and $r_0 = 33.9\,\mathrm{nm}$, yielding $A_1 = 1.20 \times 10^{-5}$.
Results for thin films with thicknesses of $1\,\mathrm{nm}$ and $8\,\mathrm{nm}$ ($\epsilon = 0.029,\,0.236$) are shown in figure \ref{fig:wavenumber}($a$). The Stokes model and lubrication model show excellent agreement, consistent with the thin-film assumption. Ultra-thin films exhibit larger wavenumbers and growth rates, primarily due to vdW forces. 
In addition, \cite{tomo2022} reported results for thicker films, up to $10\,\mathrm{nm}$ in a nanoscroll of radius $r_0 = 13.0\,\mathrm{nm}$, corresponding to $\epsilon = 0.77$, where the lubrication model may become inadequate. Figure \ref{fig:wavenumber}($b$) presents results for $\epsilon = 0.6$ and $0.7$, revealing significant discrepancies between models. This divergence arises because the lubrication model assumes $\lambda \gg \epsilon$, effectively reducing the flow to one-dimensional axial motion $w$, while thicker films exhibit substantial radial flow $u$ that must be resolved using the two-dimensional Stokes formulation. Specifically, the lubrication model overestimates both the growth rate $\omega$ and the dominant wavenumber $k_{max}$ (corresponding to the maximum growth rate, as marked in figure \ref{fig:wavenumber}$a$). Our lubrication formulation shows better agreement with the Stokes model than that of \cite{tomo2022}, primarily due to the incorporation of more accurate interfacial curvature conditions.
Notably, the cutoff wavenumber at which $\omega = 0$ remains consistent across models, since the expressions $[k^2-1/\alpha^2-3A_1/(1-\alpha)^4-3A_2/8\alpha^4]$ in (\ref{eq_dispersion_stokes}), (\ref{eq_le_dis}) and (\ref{eq_lub_tomo}) are identical, which is a consequence of the shared interfacial pressure relation given in (\ref{eq_model8}).

\begin{figure}
    \centering
    \includegraphics[width=1\linewidth]{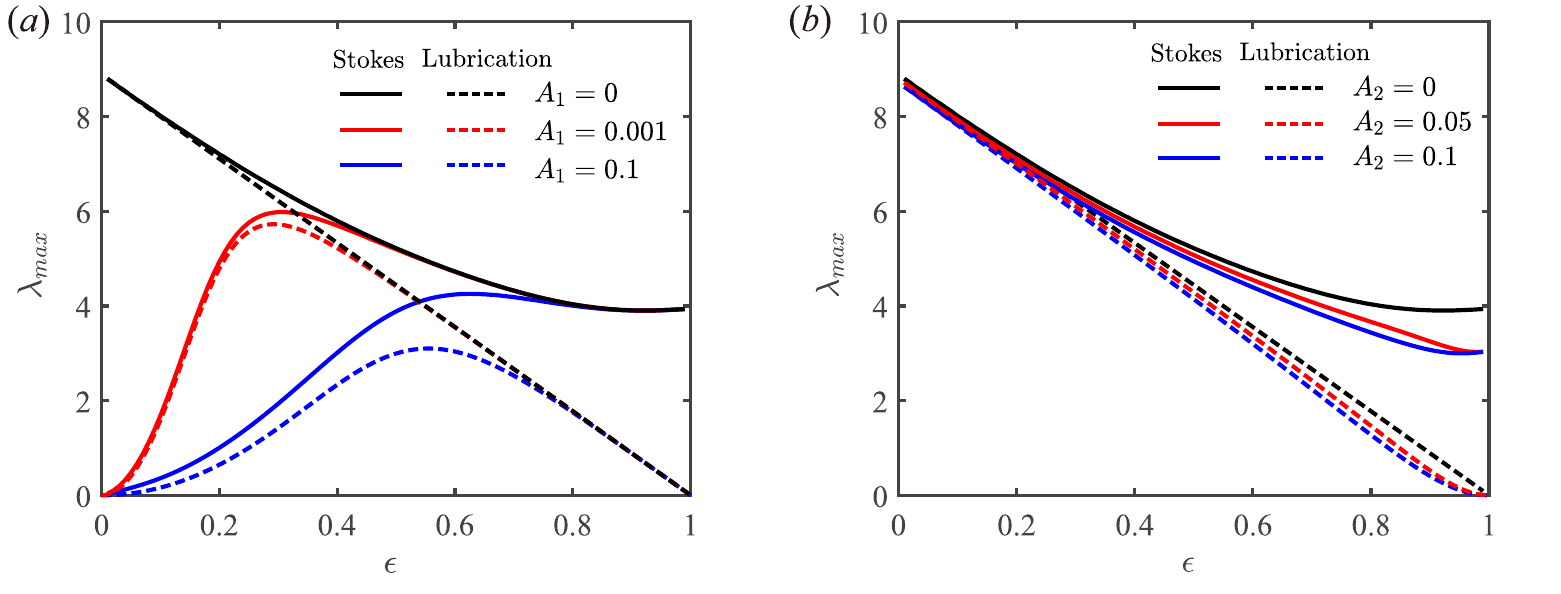}
    \caption{Influence of film thickness $\epsilon$ on the dominant wavelength $\lambda_{max}$: ($a$) different $A_1$ with $A_2=0$, ($b$) different $A_2$ with $A_1=0$. The solid lines are the predictions of Stokes model (\ref{eq_dispersion_stokes}), and the dashed lines represent the predictions of lubrication model (\ref{eq_le_dis}).}
    \label{fig:wavelength-e}
\end{figure}

These dispersion relations not only quantify the influence of vdW forces on the perturbation growth rate, but also enable determination of the dominant wavelength $\lambda_{max} = 2\pi/k_{max}$ through identification of the peak, thereby revealing the underlying physical mechanism. Figure \ref{fig:wavelength-e} illustrates the dependence of $\lambda_{max}$ on the initial film thickness $\epsilon$ for different vdW force strengths. 
In the thick-film regime, the Stokes model predicts larger dominant wavelengths than the lubrication model, where the latter gives $\lambda_{max} \approx 2\sqrt{2}\pi(1 - \epsilon)$. Figure \ref{fig:wavelength-e}($a$) demonstrates the significant influence of solid-liquid vdW forces ($A_1$): for sufficiently thin films, $\lambda_{max}$ decreases from the classical value $2\sqrt{2}\pi$ \citep{Gennes2004} towards zero. 
This phenomenon has been explained by \citet{vrij1966}, who analyzed the vdW-force-driven instability of planar liquid films. In his work, the wavelength of the fastest-growing mode is $\lambda_{max} = \epsilon^2 \sqrt{8\pi^3 \gamma / A_1}$, which approaches zero as $\epsilon \to 0$.
In the ultra-thin limit, wall curvature effects become negligible, resulting in substantial wavelength reduction—a behavior consistent with experimental findings reported by \cite{tomo2022}.
Figure \ref{fig:wavelength-e}($b$) further shows that liquid-liquid vdW forces ($A_2$) also reduce $\lambda_{max}$, particularly in thicker films where reduced interfacial distance enhances the effect of intermolecular forces.

\section{Numerical settings}
\label{sec:simulation}

The influence of vdW forces on liquid film instability in tubular geometries is investigated through direct numerical simulations in this section. The NS equations are solved using the finite element method implemented in COMSOL Multiphysics, with the arbitrary Lagrangian–Eulerian (ALE) technique employed to accurately track the sharp liquid--gas interface. In this approach, interface nodes move in a Lagrangian manner, while interior nodes follow a predefined motion strategy to update the computational domain. The ALE method combines the Lagrangian advantage of precise interface tracking with the numerical robustness of Eulerian methods by controlling mesh deformation, making it widely adopted in free-surface flow simulations such as jet breakup \citep{Martinez2020,chen2024}, film retraction \citep{marco2022}, and interfacial instability on planar substrates \citep{Moreno-Boza2020,ubal2014}.
However, the ALE method is unsuitable for simulating topological changes, such as complete processes of film rupture or collapse where fluid domains merge or separate, due to its requirement of maintaining consistent mesh connectivity throughout the simulation. Consequently, the present study focuses on the liquid film evolution up to the singularity formation.

\begin{figure}
    \centering
    \includegraphics[width=1\linewidth]{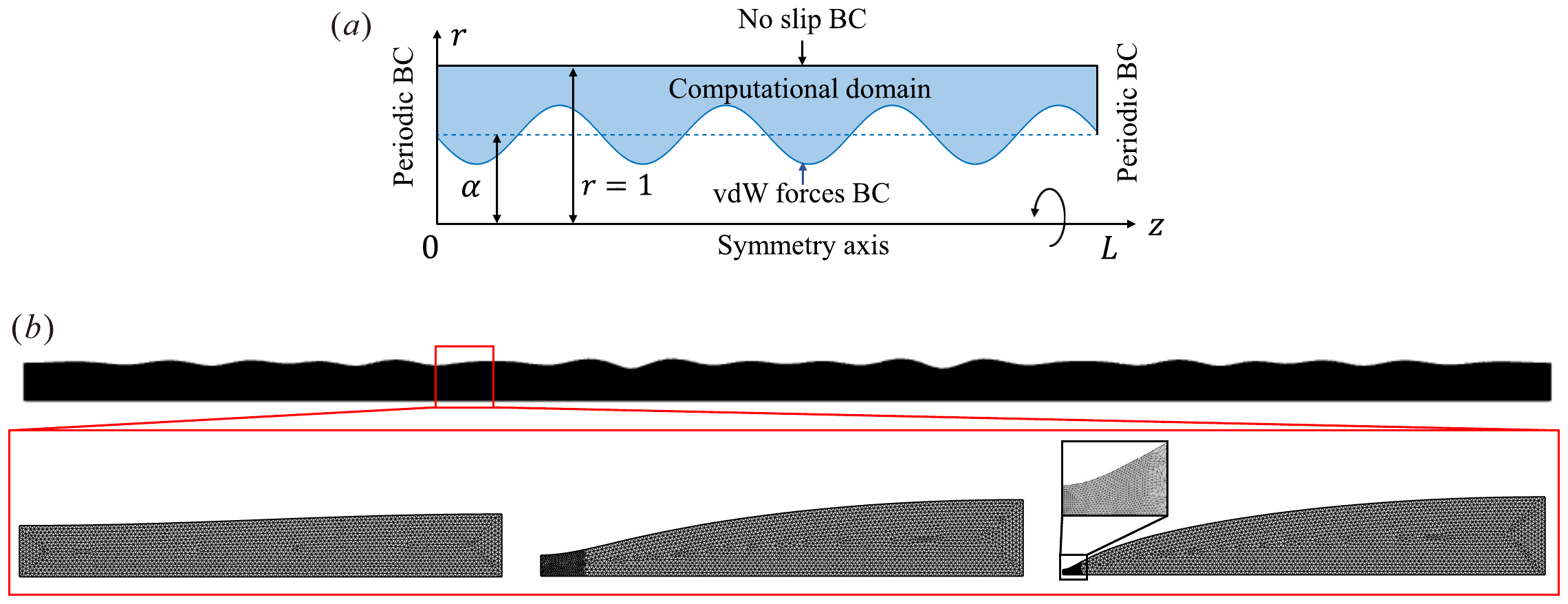}
    \caption{($a$) Quadrilateral computational domains with different boundary conditions (BCs). ($b$) Sample deformed triangular meshes for $A_1=10^{-3},\epsilon=0.1$ and three different time constants in the panels.}
    \label{fig:computation domain}
\end{figure}

The initial computational domain is a quadrilateral region (representing the cross-section of a hollow tube in cylindrical coordinates) with dimensions $[r_f, r_0] \times [0, L]$, as illustrated in figure \ref{fig:computation domain}($a$). Here, $L$ denotes the length of both the liquid film and the tube. Small perturbation $\hat{h}$ is introduced at the liquid–gas interface. Periodic boundary conditions are applied to the left and right boundaries of the domain. The top boundary satisfies a no-slip condition according to (\ref{eq_model9}), and the bottom boundary is treated as a free surface incorporating disjoining pressure as modeled by (\ref{eq_model8}).
Material properties are selected to give $\ohn =\mu / \sqrt{\rho \gamma r_0}= 10$, corresponding to a flow regime where viscous forces and surface tension dominate over inertial effects, typical of flows at the nanoscale. 
It is based on the liquid properties reported in recent experiments for cylindrical films by \cite{haefner2015} and \cite{tomo2018, tomo2022}, where $\mu = 10^{-3}\text{--}1\,\mathrm{Pa \, s}$, $\rho = 10^3\,\mathrm{kg\,m^{-3}}$, $\gamma = 0.03\text{--}0.07\,\mathrm{N\,m^{-1}}$, and $r_0 = 0.01\text{--}10\,\mathrm{\upmu m}$.
Temporal integration of the NS equations is performed using a variable-order backward differentiation formula.
The computational mesh for the liquid domain consists of non-uniform triangular elements with Lagrange basis functions, as shown in figure \ref{fig:computation domain}($b$). As the solution approaches the singularity associated with film rupture or collapse, mesh quality deteriorates in the deforming regions. To maintain accuracy, remeshing is applied near the singular region when mesh quality falls below a specified threshold (from left to right in figure \ref{fig:computation domain}$b$), enabling proper resolution of the resulting non-uniform velocity profiles. The time step is adaptively reduced during these critical phases to preserve numerical stability and accuracy.

\section{Two types of evolution: rupture and collapse}
\label{sec:two mode}

To provide a comprehensive visualization of the overall effect of vdW forces, we simulate the evolution of an extended liquid film approximately ten times longer than the dominant wavelength under random initial perturbations. The system is initialized with small-amplitude random perturbations described by $h(z)=\alpha+0.001N(z)$, where $N(z)$ follows a normal distribution with zero mean and unit variance, representing the arbitrary perturbations commonly encountered in practical situations.

\begin{figure}
    \centering
    \includegraphics[width=1\linewidth]{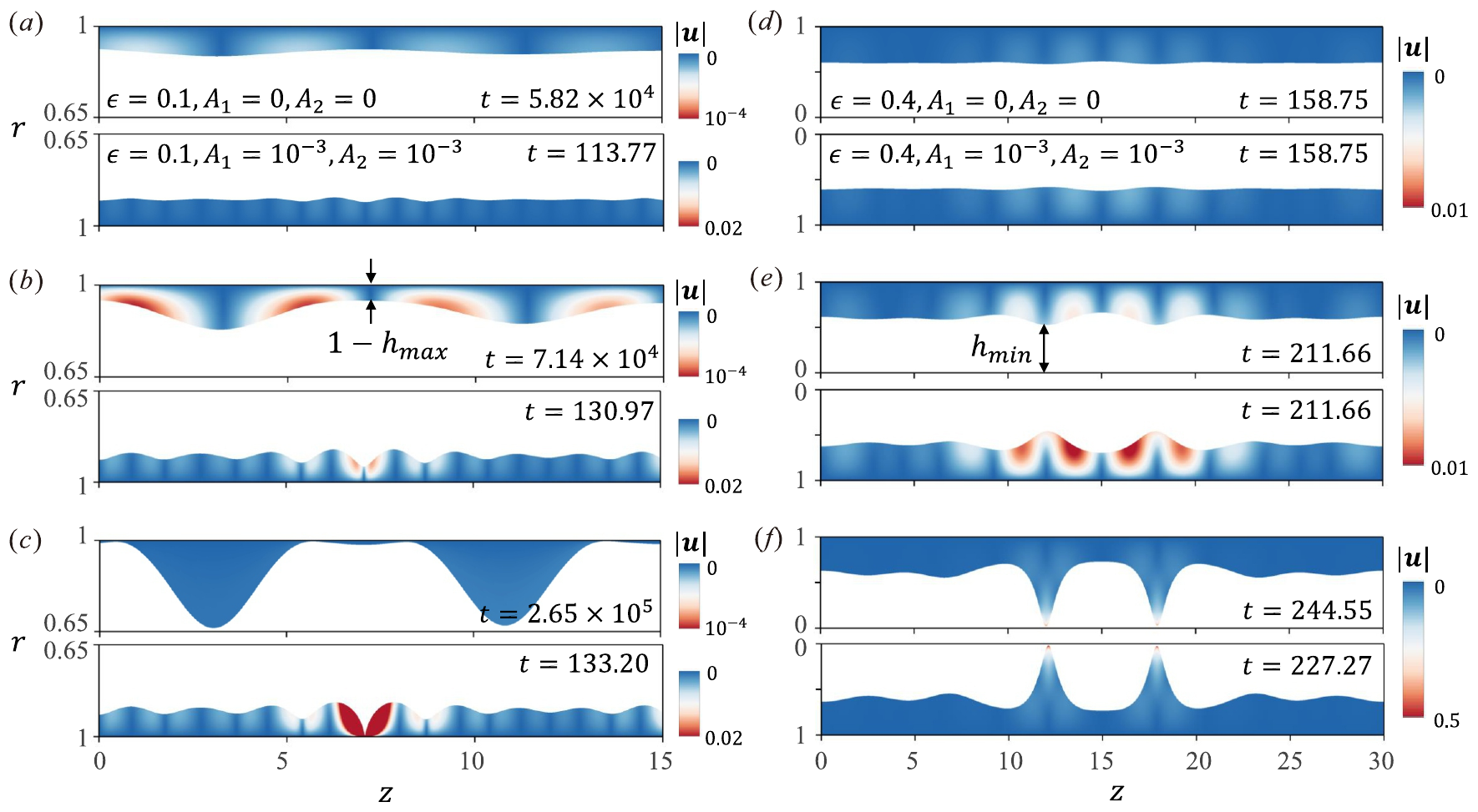}
    \caption{Evolution of a liquid film under different vdW forces, where the film is initialized as a random perturbation with a small amplitude, for ($a\text{--}c$) $\epsilon=0.1$ and ($d\text{--}f$) $\epsilon=0.4$, with comparison between $A_1=A_2=0$ (upper contour map) and $A_1=A_2=10^{-3}$ (lower contour map). The contours represent the velocity magnitude $|\textbf{\textit{u}}|=\sqrt{u^2+w^2}$.}
    \label{fig:physics}
\end{figure}

Figure \ref{fig:physics} illustrates the evolution of liquid films for two thicknesses ($\epsilon=0.1$ and $\epsilon=0.4$). In the contour plots, the upper panels show cases without vdW forces ($A_1=A_2=0$), while the lower panels correspond to cases with vdW forces ($A_1=A_2=10^{-3}$), a representative value of thin liquid films within nanotubes. Throughout the evolution process, the liquid film first develops periodic capillary waves. As perturbations grow, these waves may either rupture at the tube wall or collapse towards the central axis.
For the thinner film ($\epsilon=0.1$, figures \ref{fig:physics}$a\text{--}c$), the absence of vdW forces leads to the formation of periodic collar structures, consistent with classical Rayleigh--Plateau instability of thin films. When vdW forces are present, the wavelength is markedly reduced, and extremely high velocities develop, ultimately leading to rupture (figure \ref{fig:physics}$c$).
For the thicker film ($\epsilon=0.4$, figures \ref{fig:physics}$d\text{--}f$), collapse initiates at the film's thickest region, where vdW forces accelerate the overall evolution process.

Based on these observations, film rupture is mainly governed by vdW forces from the solid–liquid interface, whereas film collapse is primarily influenced by vdW forces through the liquid-liquid interaction. To clarify the distinct roles of vdW forces in liquid film evolution, this study separately examines the parametric effects of $A_1$ and $A_2$. When one parameter is analyzed, the other is set to zero.

\begin{figure}
    \centering
    \includegraphics[width=1\linewidth]{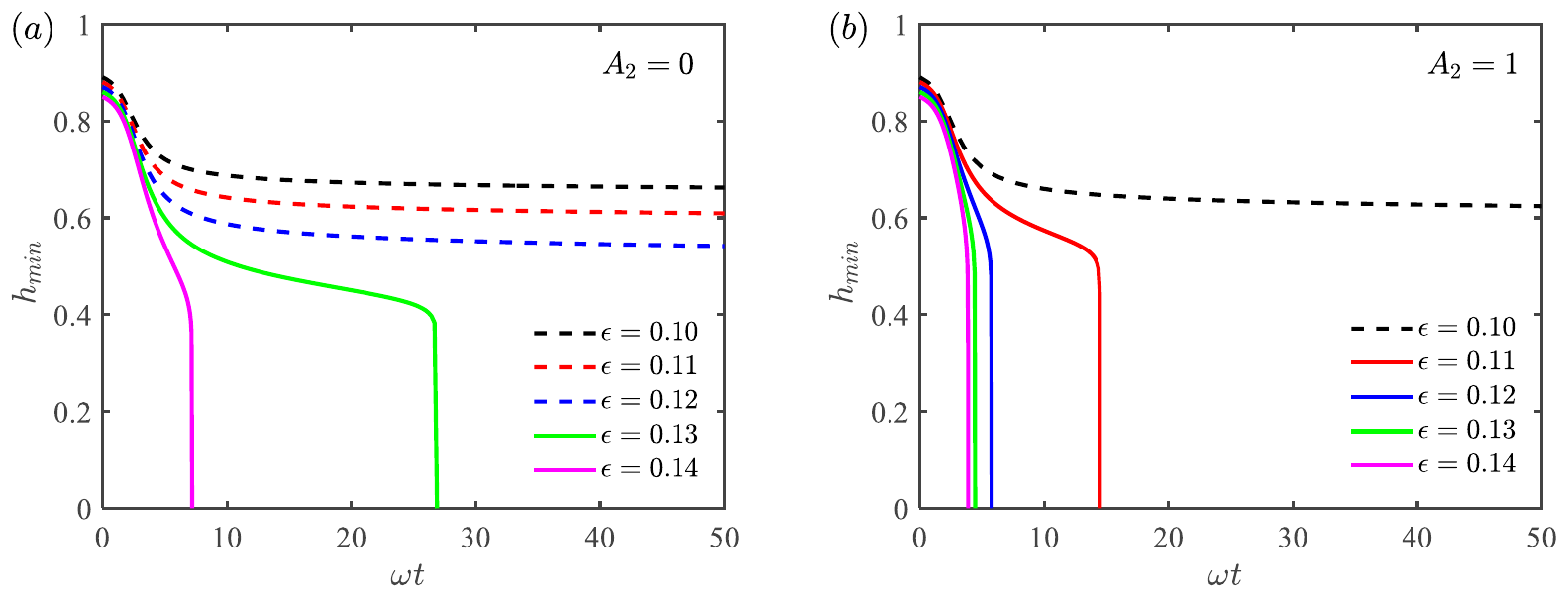}
    \caption{Evolution of the minimum radii of films $h_{ min}(t)$ for different $\epsilon$ while $A_1=0$. ($a$) $A_2=0$, ($b$) $A_2=1$. Dashed lines indicate the film remaining stable collars and solid lines indicate the film forming a plug. Time scaling $\omega t$ follows (\ref{eq_dispersion_stokes}).}
    \label{fig:different collapse}
\end{figure}

It has been observed that thicker liquid films collapse through axial coalescence to form liquid plugs, while thinner films do not. This behavior was previously reported by \cite{everett1972} and \cite{Gauglitz1988}, who defined a critical thickness $\epsilon_c = 0.12$ to determine whether a film will collapse.
To examine whether vdW forces affect this critical behavior, we simulate liquid films within a single wavelength. The initial interface profile is defined as $h_0= \alpha-0.01\cos(kz)$ with $z\in[-\pi/k,\pi/k]$, where $k$ represents the dominant wavenumber obtained from the Stokes model (\ref{eq_dispersion_stokes}).
Figure \ref{fig:different collapse} shows the temporal evolution of the minimum thickness $h_{{min}}$ for various initial film thicknesses under two conditions: $A_2 = 0$ and $A_2 = 1$. To enable comparison on a unified timescale and emphasize late-stage differences, the horizontal axis is normalized by the linear growth rate $\omega$ from (\ref{eq_dispersion_stokes}).
As shown in figure \ref{fig:different collapse}($a$), for films with $\epsilon > 0.12$, $h_{min}$ undergoes a rapid decrease starting from approximately $h_{min} = 0.4$, indicating collapse. In contrast, films with $\epsilon \leq 0.12$ transition to slow deformation and do not collapse within extended simulation times. The phenomenon occurs because the driving surface tension is dissipated by increasing viscous resistance near the wall, while the interfacial pressure gradient becomes insufficient to sustain further radial evolution. These findings align with earlier conclusions reported by \cite{Lister2006} and \cite{Dietze2015}.
However, when vdW forces between liquid interfaces are present ($A_2 = 1$), figure \ref{fig:different collapse}($b$) shows that films with $\epsilon = 0.11$ and $0.12$ shift from non-collapsing to collapsing behavior, indicating a reduction in the critical collapse thickness $\epsilon_c$.

To derive the theoretical solution for $\epsilon_c$, we incorporate the vdW force term into the model developed by \cite{Gauglitz1988}. This model assumes uniform interfacial pressure, derived as a quasi-static simplification of (\ref{eq_model8}), expressed as
\begin{equation}
    p=\frac{h''}{(1+h'^2)^{3/2}}-\frac{1}{h\sqrt{1+h'^2}} +\frac{A_1}{(1-h)^3} -\frac{A_2}{(2h)^3}=const,
    \label{eq:pconst}
\end{equation}
The resulting governing equation is an ordinary differential equation for $h$, where the prime denotes differentiation with respect to $z$.
The left side represents the total pressure. The first two terms on the right correspond to the Laplace pressure induced by interface curvature, and the last two terms describe the disjoining pressure arising from vdW forces.
Assuming collar symmetry and zero film thickness at $z=0$, the boundary conditions are specified as $h(0)=1$ and $h'(0)=0$, with $h''(0)$ remaining an undetermined parameter. Unfortunately, non-zero $A_1$ leads to numerical divergence without appropriate initial guesses. Since collapse dynamics are primarily governed by interactions between opposing liquid-gas interfaces, particular emphasis is placed on the role of liquid–liquid vdW forces ($A_2$) in determining the critical collapse regime.
Solutions to  (\ref{eq:pconst}) under varying initial conditions yield distinct equilibrium interface profiles and corresponding liquid film volumes. The result reveals the existence of a maximum film volume $V_{max}$ for each vdW forces. Detailed analysis of the film profile and volume are provided in Appendix \ref{appC}. Notably, increasing $A_2$ reduces $V_{max}$, indicating that a smaller liquid volume can be sustained in stable equilibrium under stronger liquid-liquid attractions.

The initial liquid film is treated as a hollow cylinder with cross-sectional area $\pi\bigl[1 - (1 - \epsilon)^2\bigr]$ and dominant wavelength $\lambda_{max}(\epsilon,A_2)$ determined by (\ref{eq_dispersion_stokes}). Due to volume conservation within a single wavelength, the liquid volume at the critical state equals $V_{max}$, which gives the theoretical critical thickness $\epsilon_c$ through the relation
\begin{equation}
    V_{max}=\pi[1-(1-\epsilon_{c})^2] \lambda_{max}(\epsilon_{c},A_2).
    \label{eq:critical volume}
\end{equation}
\begin{figure}
    \centering
    \includegraphics[width=0.7\linewidth]{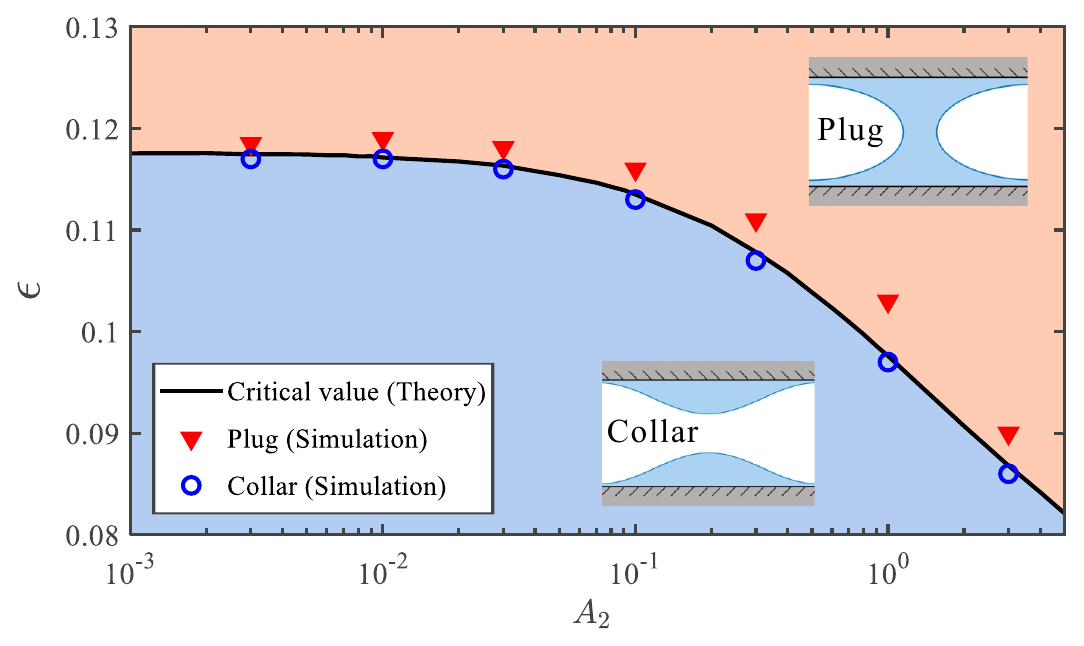}
    \caption{The critical collapse thickness $\epsilon_c$ varies with the strength of liquid–liquid vdW forces $A_2$ when $A_1=0$. The theoretical predictions from (\ref{eq:pconst})--(\ref{eq:critical volume}) are shown as a black curve, while simulation results indicating the upper and lower bounds of collar or plug are represented by symbols.}
    \label{fig:critical thickness}
\end{figure}
Using this theoretical framework, we construct a phase diagram that predicts the collapse behavior of liquid films, as shown in figure \ref{fig:critical thickness}. Collapse occurs in the red region above the theoretical curve (black solid line), while non-collapsing films correspond to the blue region below the curve.
The vdW forces between liquid--gas interfaces ($A_2$) enhance interfacial attraction towards the centre axis, thereby reducing the critical thickness $\epsilon_c$ required for collapse. To validate the theoretical predictions, lower bounds for collapse and upper bounds for non-collapse, respectively, are determined through simulations conducted within finite time scales. 
The numerical results show good agreement with theory in figure \ref{fig:critical thickness}.  
The slight overestimation in simulations arises from late-stage interface evolution. At this stage, the profile deviates from the idealized single-wave shape in (\ref{eq:pconst}), retaining volume in satellite lobes.
Additionally, when multiple waves coexist axially, deviations from $\lambda_{max}$ occur and volume is exchanged between collars. These mechanisms may collectively reduce the effective critical thickness $\epsilon_c$, a practical effect previously noted by \cite{rykner2024}.

\section{Interface evolution at the linear stage}
\label{sec:linear}

After describing the overall evolution characteristics of the liquid film, we quantitatively investigate the dynamics at each regime. When perturbation amplitudes are small, the interface morphology is approximately sinusoidal, termed the linear stage. 
Building on linear instability analysis and numerical verification via simulations, this section examines how the liquid film evolves during the linear stage.
The influence of vdW forces on the dominant wavelength is examined in \S\,\ref{sec:wavelength}, and their effect on perturbation growth rate is discussed in \S\,\ref{sec:perturbation growth}.

\subsection{Dominant wavelength}
\label{sec:wavelength}

\begin{figure}
    \centering
    \includegraphics[width=0.9\linewidth]{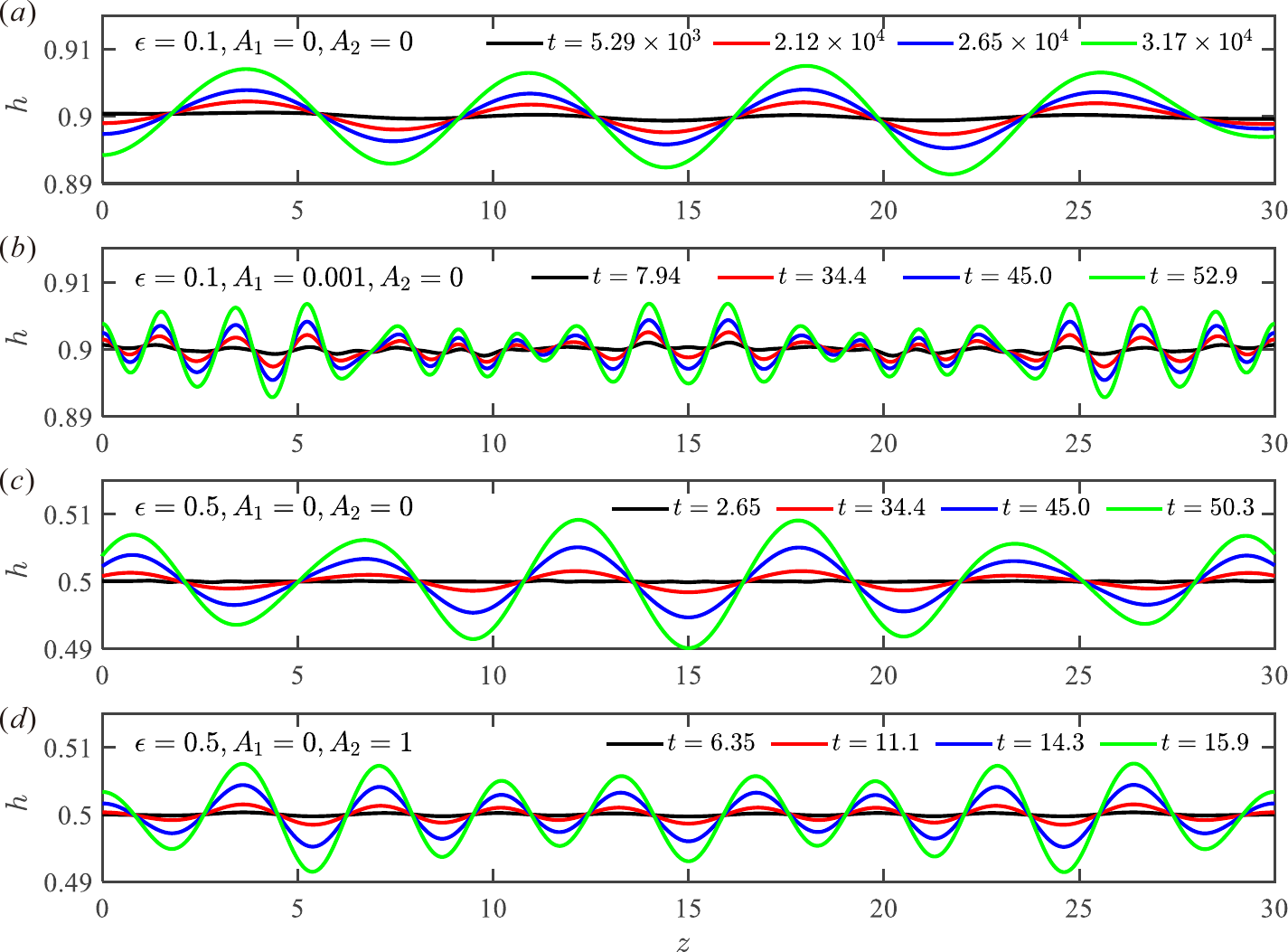}
    \caption{Interface profiles at four time instants illustrated in different colors, with film thickness ($a$,$b$) $\epsilon=0.1$ under different $A_1$, and ($c$,$d$) $\epsilon=0.5$ under different $A_2$. }
    \label{fig:spectrum shape}
\end{figure}

\begin{figure}
    \centering
    \includegraphics[width=0.9\linewidth]{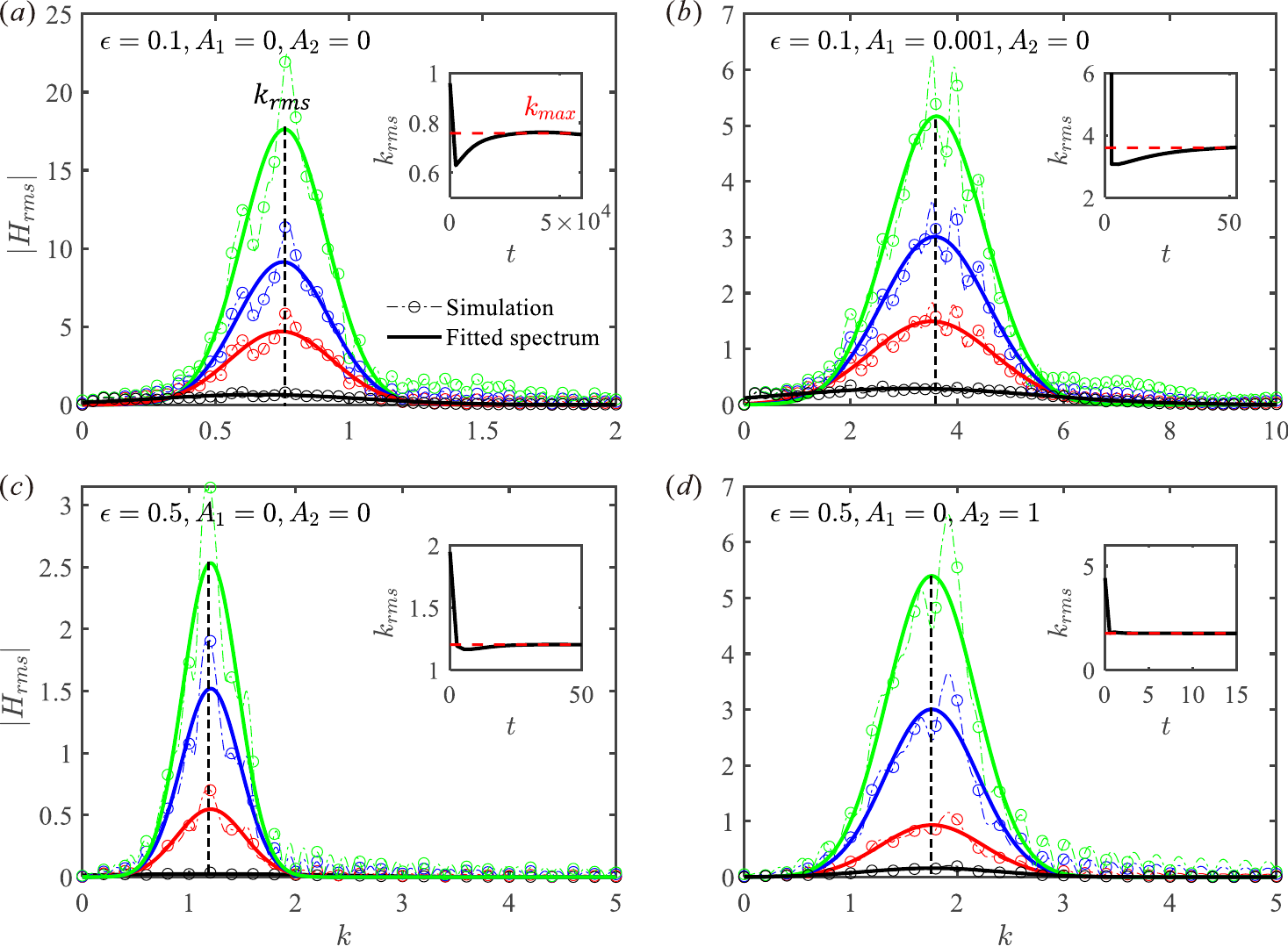}
    \caption{ The root mean square $|H_{rms}|$ of non-dimensional perturbation amplitude versus wavenumber $k$ at corresponding time instants under the same conditions as in figure\,\ref{fig:spectrum shape}. The dash-dotted lines with circles are extracted from numerical simulations fitted by the Gaussian function (solid lines). The insets show the time histories of the dominant wavenumber converging to $k_{max}$ (red dashed lines) extracted from the spectra.}
    \label{fig:spectrum}
\end{figure}
To quantitatively validate the influence of vdW forces on the dominant wavelength $\lambda_{{max}}$, we simulate liquid films with various initial thicknesses under different values of $A_1$ and $A_2$. The axial length of the computational domain is set to approximately 10 times $\lambda_{{max}}$ to sufficiently resolve the wave distribution, maintaining consistency with the simulation configuration used in figure \ref{fig:physics}.
Under the combined action of surface tension and vdW forces, the initially small random perturbations gradually develop into well-defined capillary waves. Figure \ref{fig:spectrum shape} shows the evolution of the interface profile $h(z,t)$ under four distinct parameter conditions, with the same initial perturbation form $N(z)$ used across all cases to isolate parameter effects. Both $A_1$ (figures \ref{fig:spectrum shape}$a$,$b$) and $A_2$ (figures \ref{fig:spectrum shape}$c$,$d$) accelerate the evolution rate and reduce the capillary wavelength.

To compare numerical results with theoretical predictions from the Stokes model (\ref{eq_dispersion_stokes}), we conduct multiple independent simulations (30 realisations per parameter) with different random initial conditions to statistically determine the dominant wavelength $\lambda_{{max}}$. For each realization, we use the discrete Fourier transform of the interface profile $h(z,t)$ to obtain the power spectral density (PSD) of the perturbations. This statistical methodology, introduced by \cite{zhao2019}, has been successfully applied to characterize dominant instability modes in various liquid film systems \citep{zhao2022,zhao2023}. 
Figure \ref{fig:spectrum} displays the square root of the ensemble-averaged PSD, denoted as $H_{{rms}}$, at each time step for the conditions corresponding to figure \ref{fig:spectrum shape}. A Gaussian function is fitted to the modal distribution spectrum, with the spectral peak identifying the wavenumber $k_{{rms}}$. The insets illustrate the temporal evolutions of $k_{{rms}}$ extracted from simulations, showing rapid convergence to a stable value that we define as the dominant wavenumber $k_{{max}}$.
Consistent with the observations in figure \ref{fig:spectrum shape}, the dominant wavenumbers in figures \ref{fig:spectrum}($b$,$d$) with vdW forces are significantly larger than those in figures \ref{fig:spectrum}($a$,$c$) without vdW forces, confirming the wavelength-reducing effect of the intermolecular interactions.

\begin{figure}
    \centering
    \includegraphics[width=1\linewidth]{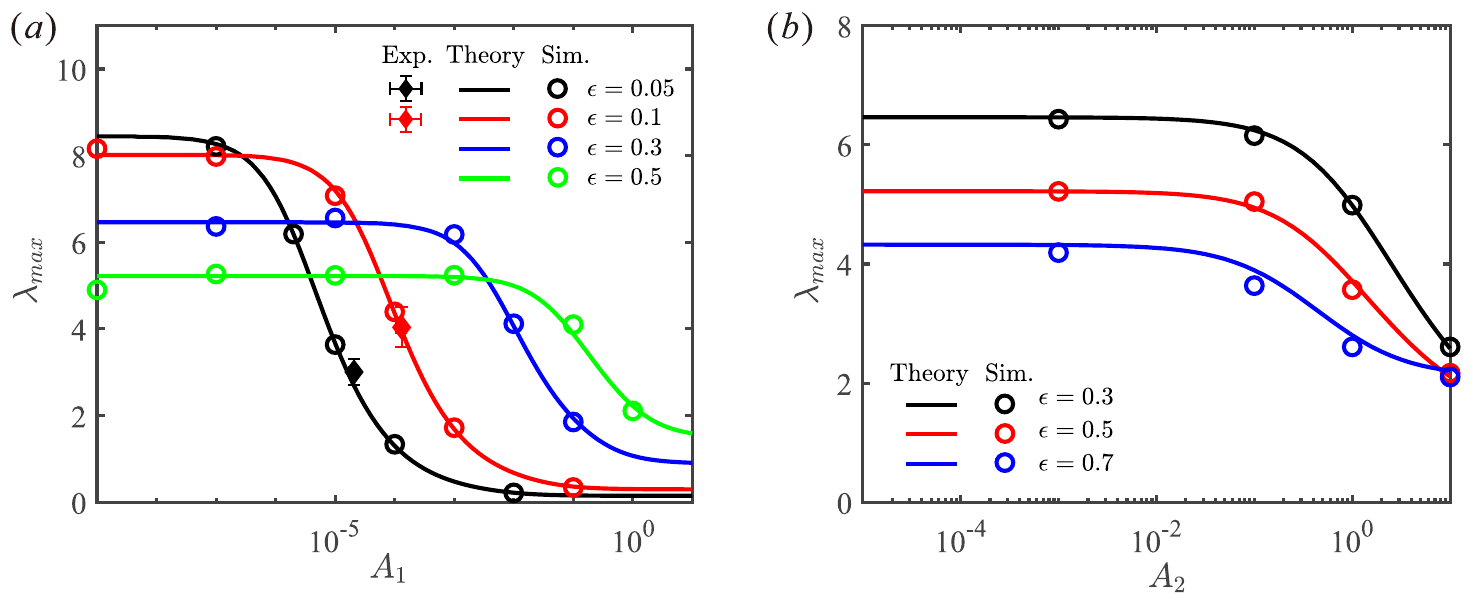}
    \caption{Effects of vdW forces ($a$) $A_1$ and ($b$) $A_2$ on the dominant wavelengths $\lambda_{ max}$: a comparison between the theoretical predictions of (\ref{eq_dispersion_stokes}) (solid lines) and results of numerical simulation (Sim.) from the spectra (hollow symbols) with different film thickness $\epsilon$. The solid diamonds represent experimental results (Exp.) by \cite{tomo2022}.}
    \label{fig:wavelength-a}
\end{figure}

Based on the preceding analysis, we present the simulated values of $\lambda_{{max}}$ for various parameter combinations in figure \ref{fig:wavelength-a}, as well as experimental data from \cite{tomo2022}.
Both numerical results and experimental measurements demonstrate excellent agreement with our theoretical model, confirming that the proposed theoretical framework accurately captures the physical mechanism of instability across a wide parameter range.
Furthermore, the results clearly show that vdW forces significantly reduce the dominant wavelength during the linear evolution stage of liquid films, with this effect being pronounced over a broad spectrum of conditions. In general, thinner films are more sensitive to solid-liquid vdW forces ($A_1$), while thicker films show earlier and more substantial influence from liquid-liquid vdW forces ($A_2$).

\subsection{Perturbation growth}
\label{sec:perturbation growth}

\begin{figure}
    \centering
    \includegraphics[width=1\linewidth]{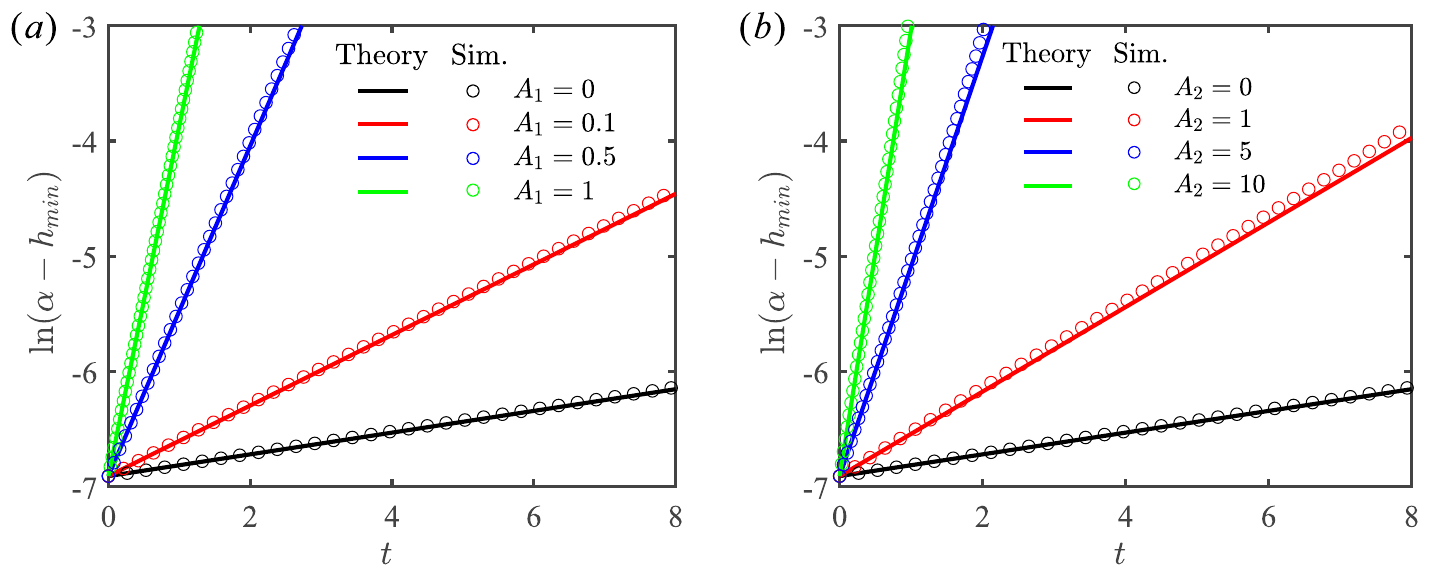}
    \caption{Linear time evolution of the minimum radii of films $h_{ min}$ with various ($a$) $A_1$, where the maximum growth rate is $\omega=0.094,0.306,1.433,3.041$, respectively; ($b$) $A_2$, for $\omega=0.094,0.367,1.823,3.879$, respectively. The numerical solutions (circles) are compared with the predictions of the Stokes model (solid lines). Here, $\epsilon=0.5$, $\hat{h}=10^{-3}$.}
    \label{fig:growth rate}
\end{figure}

During the linear stage of liquid film evolution, the interface profile under the small perturbation assumption follows $h(z,t) = \alpha + \hat{h}\eu^{\omega t+\iu kz}$, where $\hat{h}$ represents the initial perturbation amplitude at $t=0$. To validate the theoretical predictions, we simulated a perturbed liquid film within one dominant wavelength, initializing the interface as $h_0 = \alpha - 10^{-3} \rm{cos}(\it kz)$. The early-stage evolution results, presented in figure \ref{fig:growth rate}, demonstrate excellent agreement with the Stokes model (\ref{eq_dispersion_stokes}). Additionally, the simulations confirm that vdW forces accelerate perturbation growth during the linear instability development of liquid films.

\begin{figure}
    \centering
    \includegraphics[width=1\linewidth]{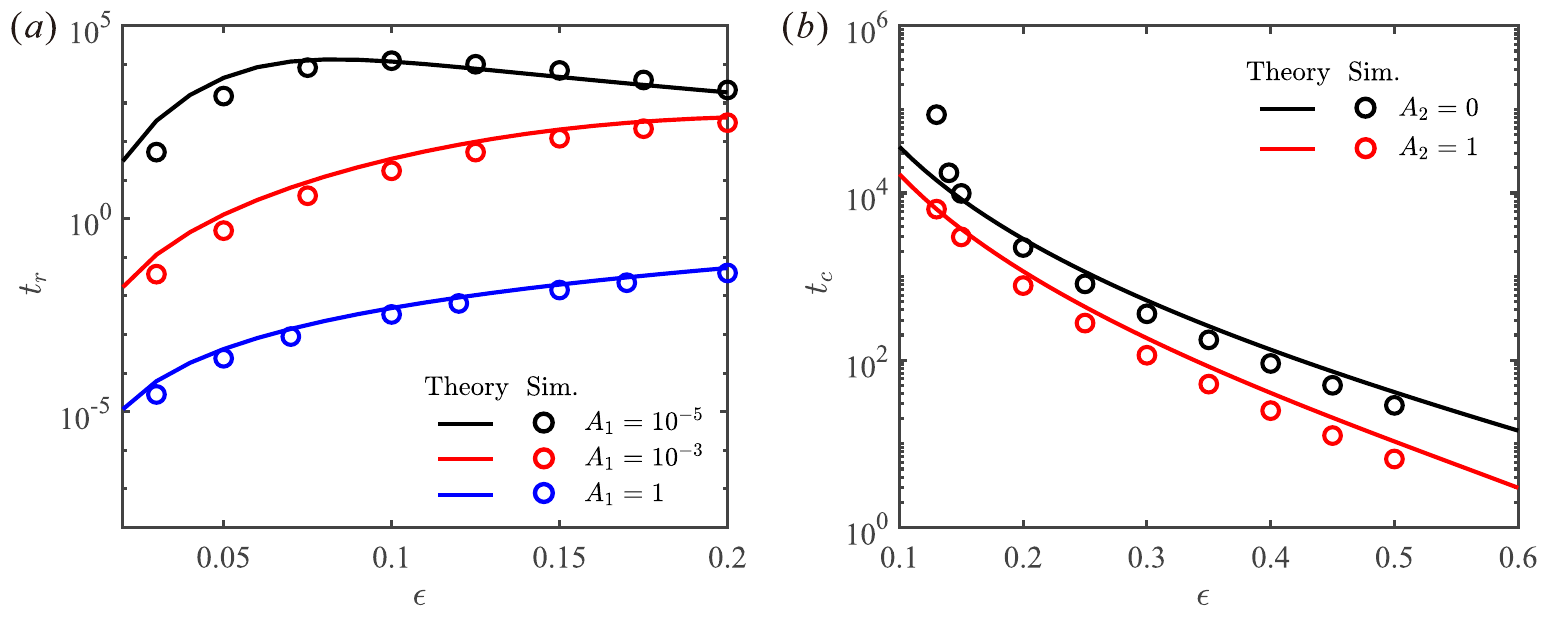}
    \caption{Linear prediction of the ($a$) rupture time $t_r$ and ($b$) collapse time $t_c$. The numerical solutions (symbols) are compared with the predictions of the Stokes model (solid lines).}
    \label{fig:singular time}
\end{figure}

Furthermore, predicting the time to reach singularity is valuable for both computational modeling and practical applications. These times are denoted as $t_r$ for rupture and $t_c$ for collapse in liquid films under vdW forces. 
We estimate the singularity time based on the linear stage, which is easily predicted and usually accounts for the vast majority. This method has been previously applied to film rupture on planar substrates and film collapse \citep{Martinez2021, rykner2024}. The specific formulas used are $t_r=[\ln (\epsilon/\hat{h})]/\omega\,,\,t_c=[\ln (1-\epsilon)/\hat{h}]/\omega$, where $\omega$ is the corresponding linear growth rate according to (\ref{eq_dispersion_stokes}).
In our simulations, the initial interface configuration is the same as the case used for growth rate validation.
Figure \ref{fig:singular time} compares the simulated rupture and collapse times with theoretical predictions. Overall, the linear growth rate based theory captures both the trend and magnitude of the evolution duration. Regarding the role of vdW forces, solid–liquid interactions ($A_1$) reduce the rupture time, as shown in figure \ref{fig:singular time}($a$). 
For small $A_1$, thinner films do not rupture more rapidly due to their weak curvature effects, which limit the growth rate. In contrast, large $A_1$ values significantly accelerate film evolution and promote earlier rupture.
Meanwhile, liquid–liquid interfacial interactions ($A_2$) consistently shorten the collapse time across all film thicknesses, as is intuitively expected and demonstrated in figure \ref{fig:singular time}($b$).
Theoretical predictions slightly overestimate the simulation results generally, as interface evolution accelerates in the nonlinear regime relative to the linear growth rate. 
However, near the critical collapse thickness (figure \ref{fig:singular time}$b$), simulations show significant deviations. This discrepancy stems from a viscous blocking phenomenon—identified by \cite{Dietze2015}—in which the thin liquid film between adjacent collars strongly inhibits further evolution. Consequently, the film remains in a prolonged quasi-stable state during the nonlinear stage before undergoing rapid collapse, as seen in the late-time evolution shown in figure \ref{fig:different collapse}.

\section{Dynamics at the nonlinear stage}
\label{sec:nonlinear}

As introduced in \S\,\ref{sec:linear}, linear instability analysis is based on the assumption of small-amplitude perturbations, where the evolving interface morphology can be represented as a superposition of sinusoidal Fourier modes. However, as perturbations grow into the nonlinear regime, the interface profile progressively deviates from simple trigonometric forms. This section investigates the nonlinear stages of interface evolution under vdW forces, examining the morphology and structure immediately preceding rupture in \S\,\ref{sec:rupture} and collapse in \S\,\ref{sec:collapse}, as well as the self-similar behavior in \S\,\ref{sec:self-similarity} during these processes.

\subsection{Film rupture and satellite lobes}
\label{sec:rupture}

\begin{figure}
    \centering
    \includegraphics[width=1\linewidth]{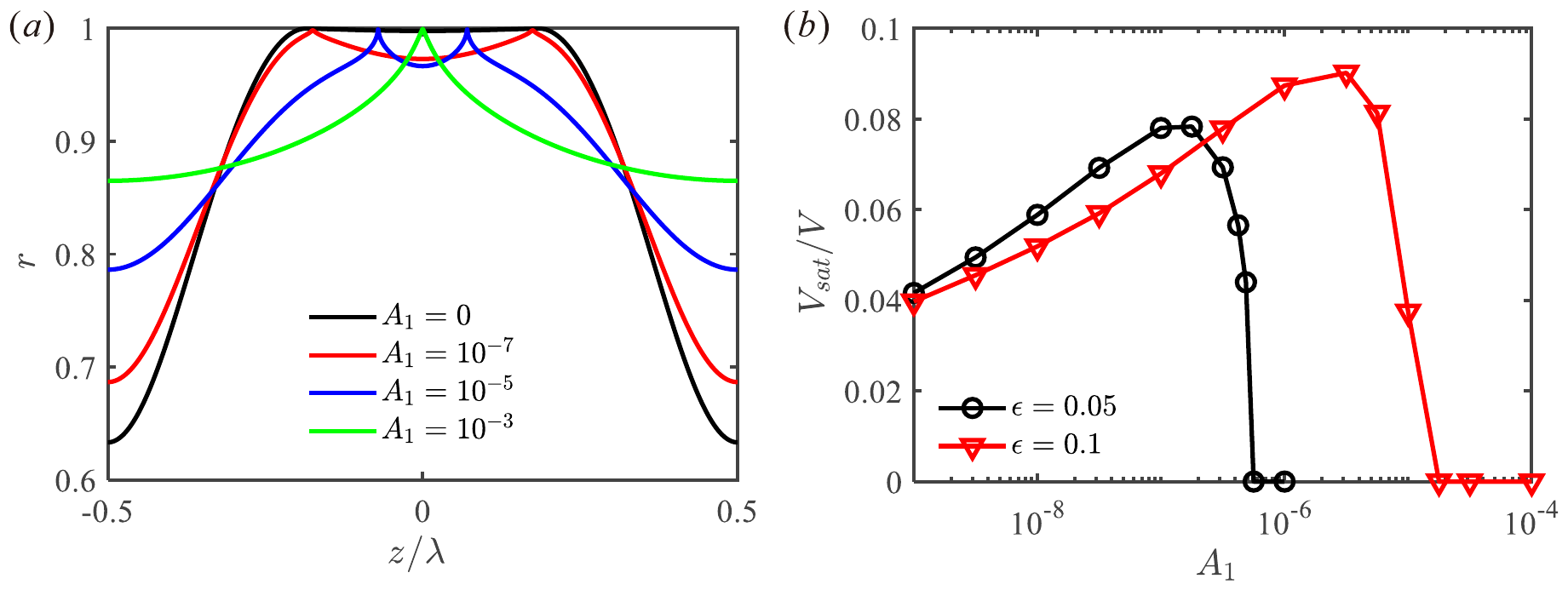}
    \caption{($a$) The film profile near rupture with different $A_1$ for $\epsilon=0.1$. ($b$) Volume proportion of satellite lobes under the influence of $A_1$. }
    \label{fig:satellite size}
\end{figure}

To examine the nonlinear rupture behavior of liquid films in detail, we construct a computational domain spanning one wavelength. The initial mean thickness is set to $\epsilon=0.1$, below the critical collapse thickness $\epsilon_c$ to ensure that rupture occurs. A sinusoidal initial profile with amplitude 0.001 is imposed. As shown in figure \ref{fig:satellite size}($a$), vdW forces significantly influence the late-stage morphology of the liquid film.
In the absence of vdW forces, the film evolves into primary collars accompanied by satellite lobes. When vdW forces are present, large radial velocities develop at the thinnest region of the film, ultimately leading to rupture.
To consistently identify satellite lobes, we define the thinnest point as satisfying $1-h_{{max}} < 10^{-4}$, minimizing misidentification due to morphological changes from flow between primary and satellite structures.
As $A_1$ increases, the relative length of satellite lobes gradually decreases until they fully disappear. Correspondingly, the volume ratio $V_{{sat}}/V$ also varies, as shown in figure \ref{fig:satellite size}($b$), where $V_{{sat}}$ is the satellite lobe volume, and $V$ is the total liquid volume per wavelength. As $A_1$ increases, $V_{{sat}}/V$ initially rises, peaks at approximately 0.09, and then rapidly decreases to zero. Thicker films require a higher threshold of $A_1$ to completely suppress satellite lobe formation.

\begin{figure}
    \centering
    \includegraphics[width=1\linewidth]{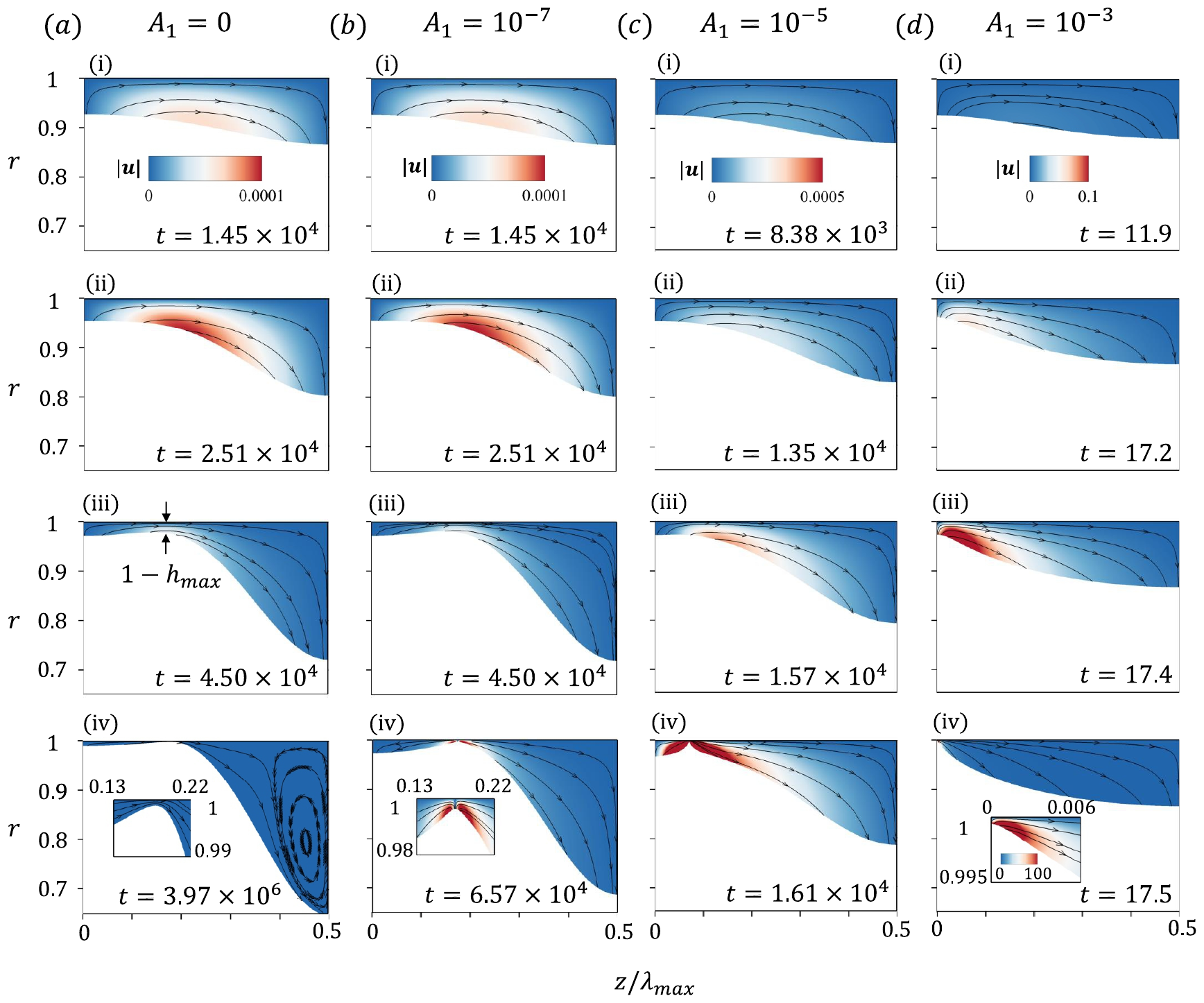}
    \caption{Evolution of perturbation growth at the nonlinear stage with four $A_1$ values: ($a$) $A_1=0$, ($b$) $A_1=10^{-7}$, ($c$) $A_1=10^{-5}$, ($d$) $A_1=10^{-3}$. The contours represent the velocity magnitude $|\textbf{\textit{u}}|=\sqrt{u^2+w^2}$, and the arrowed lines represent streamlines. Here, $\epsilon=0.1$.}
    \label{fig:satellite lobe}
\end{figure}

To further investigate the mechanism governing profile variation under vdW forces, we present velocity contours of $|\textbf{\textit{u}}| = \sqrt{u^2 + w^2}$ and corresponding streamlines at different evolution times in figure \ref{fig:satellite lobe}.
In the absence of vdW forces as shown in figure \ref{fig:satellite lobe}($a$), instability is driven solely by surface tension. The film in the initially thinnest region experiences increasing adhesive effects near the wall, leading to flow deceleration. As a result, the film between developing collars bends inwards, forming secondary satellite lobe structures \citep{Dietze2015}.
A slow axial flow from the satellite lobe towards the primary collar persists, continually reducing the satellite lobe volume.
However, vdW forces alters this evolution by introducing a molecular length scale $a = \sqrt{\tilde{A}/6\pi\gamma} = \sqrt{A}r_0$ \citep{Gennes1985}. 
When the vdW forces are weak ($A_1 = 10^{-7}$, figure \ref{fig:satellite lobe}$b$), surface tension remains the dominant driving mechanism through most of the evolution, until the film thickness approaches $a$.  
Subsequently, growing vdW attraction induces rapid rupture towards the wall at the thinnest point (figure \ref{fig:satellite lobe}$b$ iv), which suppresses axial flow towards the primary collars, and increases the relative volume of satellite lobes.
For stronger vdW forces (figure \ref{fig:satellite lobe}$c$), the region of maximum velocity (figure \ref{fig:satellite lobe}$c$ ii) shifts towards the symmetry axis, and the thinning neck contracts earlier. The resulting rupture produces a shorter satellite lobe (figure \ref{fig:satellite lobe}$c$ iv). When $A_1$ is further increased to $10^{-3}$ (figure \ref{fig:satellite lobe}$d$), vdW forces dominate over surface tension, reducing the evolution time by three orders of magnitude. The film develops strong radial velocities (figure \ref{fig:satellite lobe}$d$ iv) before any distinct satellite structure can form, maintaining the thinnest point at the centre, and fully suppressing satellite lobe development.

\begin{figure}
    \centering
    \includegraphics[width=1\linewidth]{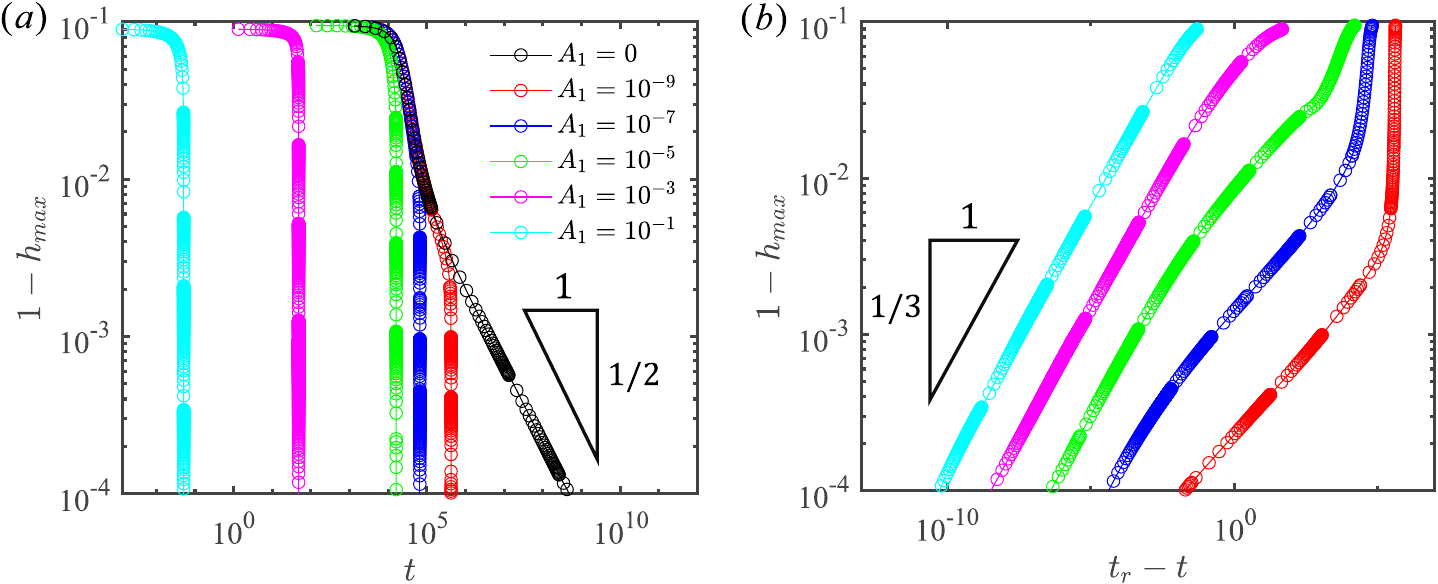}
    \caption{Minimum film thickness $1-h_{ max}$ as a function of ($a$) the time $t$ and ($b$) the time remaining to rupture $t_r-t$ for different $A_1$. The initial conditions $h_0= 0.9+10^{-3}  \rm{cos}(\it kz)$ use the corresponding dominant wave numbers.}
    \label{fig:rupture law}
\end{figure}

The nonlinear dynamics is usually evaluated by temporal evolution of the minimum film thickness $1-h_{{max}}$, with particular focus on the scaling laws governing film rupture processes \citep{zhang1999,Moreno-Boza2020,Calvo2025}. 
As shown in figure \ref{fig:rupture law}($a$), films without vdW forces thin continuously according to $1-h_{{max}} \sim t^{-1/2}$ during the nonlinear stage without rupturing. 
Capillary and viscous forces maintain the film at a finite thickness over extended durations. This behavior follows the scaling law reported by \cite{Lister2006} for annular films in the specific configuration where a collar adjoins a lobe.
When $A_1 > 0$, the thinning behavior deviates from the scaling law and leads to rapid rupture. The onset of this deviation occurs earlier as $A_1$ increases.

Based on the preceding analysis of scaling laws in liquid film rupture \citep{zhang1999,Martinez2021}, we examine the evolution on the singularity time scale $\tau = t_r - t$ (figure \ref{fig:rupture law}$b$). As $A_1$ increases, the thinning behavior progressively approaches a $\tau^{1/3}$ power law, indicating the growing dominance of vdW forces instead of surface tension.
This behavior can be explained through dimensional analysis, which constructs dimensionless relationships among the key physical quantities governing the process \citep{curtis1982}. The parametric dependence of the film thickness $\tilde{h}_f = r_0 - \tilde{h}$ can be expressed as:
\begin{equation}
    \tilde h_f =F(\tilde z, \tilde \tau, \mu, \tilde A_1),
    \label{eq:sc1}
\end{equation}
where tildes denote the dimensional versions of the flow variables.
Choosing $(\tilde\tau,\mu, \tilde A_1)$ as the dimensional basis, the Buckingham $\Pi$ theorem provides the reduced functional form:
\begin{equation}
    \tilde h_f =\tilde{H}[(\mu/\tilde A_1 \tilde \tau)^{1/3} \tilde z](\tilde A_1/\mu)^{1/3} \tilde\tau^{1/3}.
    \label{eq:sc2}
\end{equation}
This expression shows that as the rupture singularity is approached ($\tau \to 0$), the interface evolution $\tilde{h}_f$ separates into two parts: a self‑similar spatial profile $\tilde{H}$ and a time‑scaling law $\tilde{\tau}^{1/3}$. Consequently, the thickness at the rupture point ($z=0$) exhibits a $\tau$‑dependent scaling near the singularity, which leads to the dimensionless scaling relation:
\begin{equation}
    1-h_{max}\sim \tau^{1/3}\,,\,\rm when\,\tau\to 0\,.
    \label{eq:scaling}
\end{equation}
The $\tau^{1/3}$ scaling law for vdW-dominated rupture can be obtained by combining (\ref{eq:sc2}) and figure \ref{fig:rupture law}($b$), which is consistent with previous findings for thin films on planar substrates \citep{Moreno-Boza2020}.

\subsection{Film collapse}
\label{sec:collapse}
\begin{figure}
    \centering
    \includegraphics[width=0.6\linewidth]{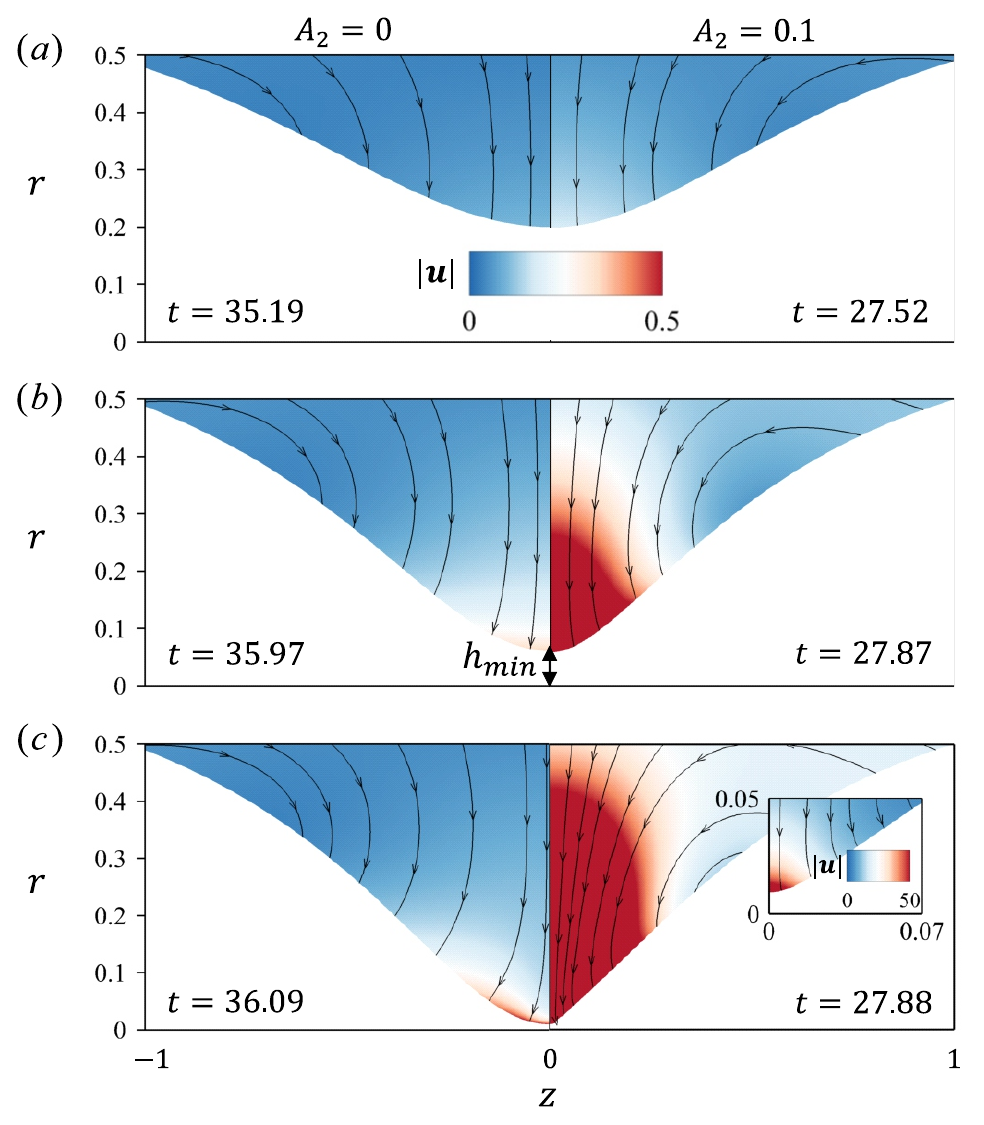}
    \caption{Evolutions of perturbation growth streamlines near film collapse for $A_2=0$ (left) and $A_2=0.1$ (right). The contours represent the velocity magnitude $|\textbf{\textit{u}}|=\sqrt{u^2+w^2}$, and the arrowed lines represent streamlines. Here, $\epsilon=0.5$.}
    \label{fig:collapse shape}
\end{figure}

Thin liquid films tend to rupture towards the tube wall, whereas thick films collapse towards the central axis. Collapse constitutes another type of singularity distinct from film rupture. This subsection examines how vdW forces affect the flow characteristics preceding liquid film collapse. The simulation setup employed an initial sinusoidal interface profile within a single wavelength, with initial thickness $\epsilon = 0.5$.
Figure \ref{fig:collapse shape} shows contours of flow velocity magnitude and corresponding streamlines at different instants. The vdW forces from the liquid–liquid interaction accelerate all stages of the collapse process, and enhance internal flow velocities within the film. 
Moreover, as shown in figure \ref{fig:collapse shape}($c$), vdW forces cause the flow to focus more strongly towards the tip region, and sharpen the interface profile gradient.
Owing to the $\Pi \propto d^{-3}$ dependence of vdW forces, they drive radially dominated flow, where the tip velocity reaches approximately twice that without vdW forces.

\begin{figure}
    \centering
    \includegraphics[width=0.7\linewidth]{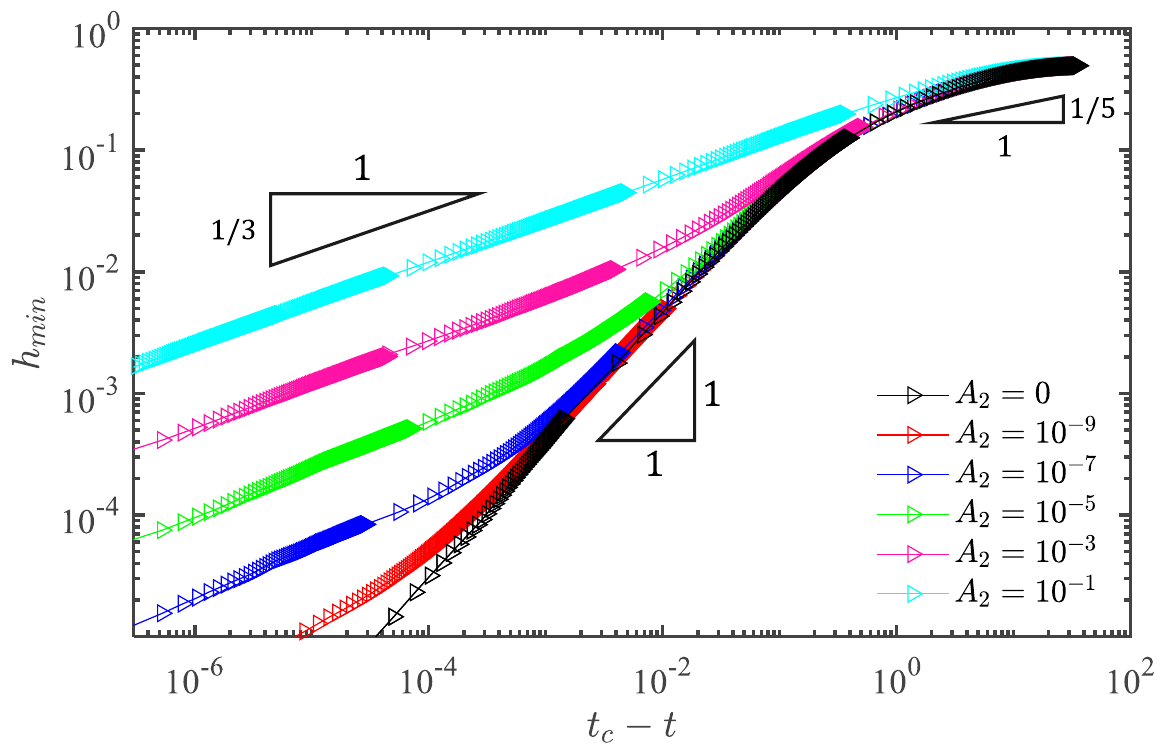}
    \caption{Minimum interface radius $h_{ min}$ as a function of time remaining to collapse $t_c-t$ for different $A_2$. The initial conditions $h_0= 0.5-10^{-3} \rm {cos}(\it kz)$ use the corresponding dominant wavenumbers. }
    \label{fig:collapse law}
\end{figure}

We further examine the nonlinear scaling behavior near the collapse singularity, focusing on the evolution of the minimum point $h_{{min}}$ with respect to the time to collapse $t_c - t$, as shown in figure \ref{fig:collapse law}.
In the absence of vdW forces, where viscous resistance and surface tension dominate, the film collapse initially follows $h_{{min}} \sim (t_c - t)^{1/5}$. During this early stage, the interface evolution could be described by the lubrication theory (\ref{eq_le_dis}) and exhibits self-similar behavior \citep{ding2019}, as the film maintains a small-amplitude morphology. This $1/5$ scaling law is also consistent with the initial stages of film collapse observed experimentally \citep{pahlavan2019}.
Subsequently, the evolution transitions to a linear scaling $h_{{min}} \sim (t_c - t)$, which aligns with the behavior reported for bubble pinch-off in highly viscous liquids, both in capillary tubes \citep{pahlavan2019} and in large tanks \citep{burton2005}, since these processes represent temporally localized singularities dominated by radial viscous effects.
In contrast, when vdW forces are present, the collapse process converges to a $1/3$ power-law scaling, identical to the rupture scaling law. This indicates that vdW forces dominate the final stages of collapse, overriding the geometric constraints that govern collapse in their absence.

\subsection{Self-similarity}
\label{sec:self-similarity}
For these singular behaviors governed by vdW forces, a unified description applicable to both rupture and collapse can be established. We rescale the system parameters using the characteristic vdW length $a=\sqrt{\tilde{A}/6\pi\gamma}$ (denoted by subscript $a$), as
\begin{equation}
    (r_a,z_a,h_a,t_a)=\frac{r_0}{a}(r,z,h,t),\quad 
    (p_a,\boldsymbol{\sigma}_a)=\frac{a}{r_0}(p,\boldsymbol{\sigma}).
    \label{eq:rescale_a}
\end{equation}

\begin{figure}
    \centering
    \includegraphics[width=1\linewidth]{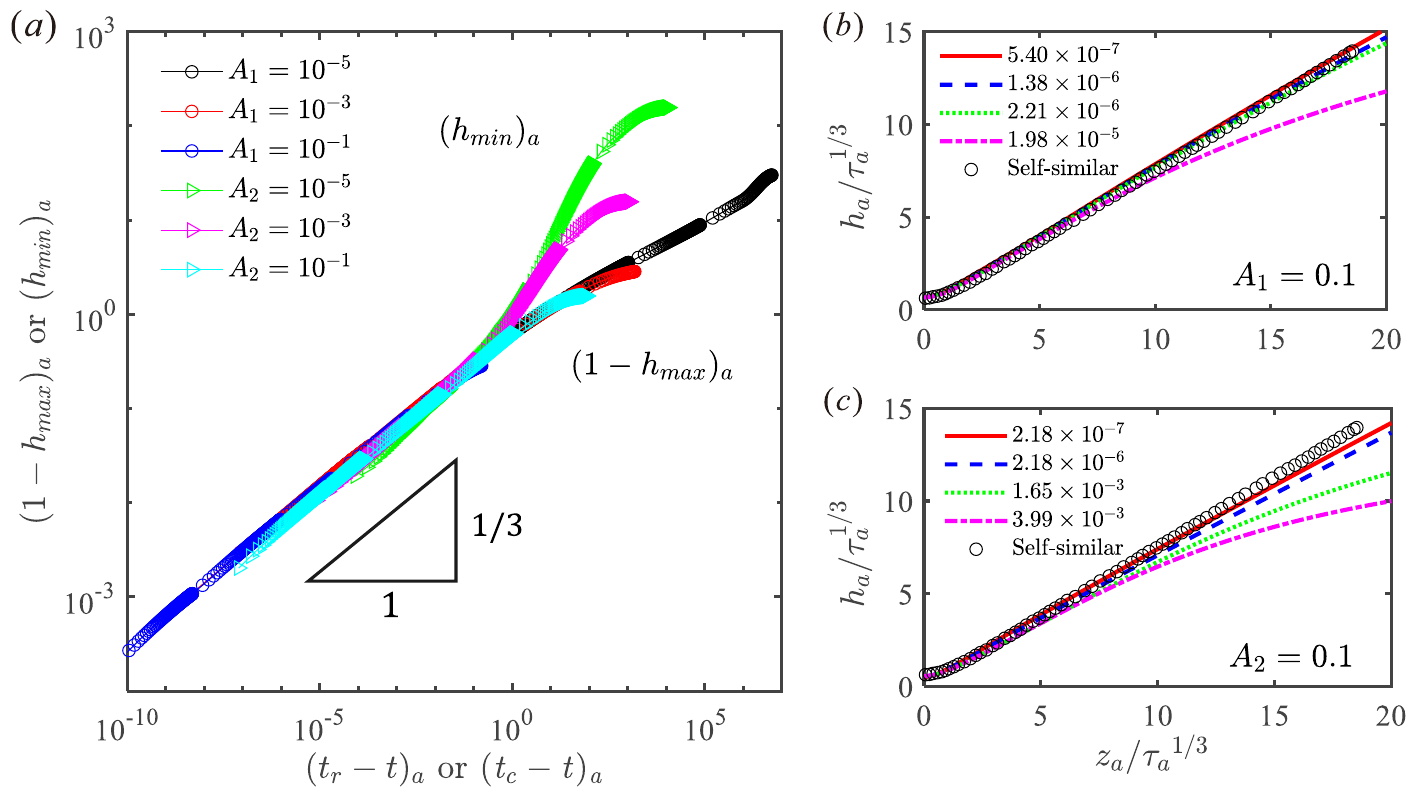}
    \caption{($a$) Extreme point of interface evolution $(1-h_{max})_a$ or $(h_{min})_a$ as a function of the time remaining to singular points $(t_r-t)_a$ or $(t_c-t)_a$. The length and time are rescaled by the characteristic length $a=\sqrt{\tilde{A_i}/6\pi\gamma}$, where $i=1,2$ for rupture and collapse respectively. ($b,c$) Self-similarity profile at different instants $\tau_a$ for $A_1=0.1$ or $A_2=0.1$, compared with theoretical solution of film on a flat substrate by \cite{Moreno-Boza2020}.}
    \label{fig:singular}
\end{figure}

Figure \ref{fig:singular}($a$) shows the scaling behavior near the singularity after this nondimensionalisation, with rupture and collapse data corresponding to figures \ref{fig:rupture law} and \ref{fig:collapse law}, respectively. All datasets ultimately collapse onto the universal curve $h_a \sim \tau_a^{1/3}$, confirming the common scaling governing both types of singular evolution.
We further hypothesize that the flow near the singularity exhibits local self-similarity, where the interface profile can be derived from (\ref{eq:sc2}), expressed as
\begin{equation}
     h_a \xrightarrow{\tau_a\to 0} H(z_a/\tau_a^{1/3})\tau_a^{1/3}.
    \label{eq:self-similar}
\end{equation}
Figures \ref{fig:singular}($b$,$c$) present the self-similar profiles for rupture and collapse under selected parameters, with legend values indicating $\tau_a$. As $\tau_a \to 0$, the interface profiles converge towards a wedge-like shape with a constant opening angle. Comparison with the self-similar solution for planar film rupture reported by \cite{Moreno-Boza2020} shows good agreement with our simulations in the cylindrical tube, supporting the universality of vdW-dominated singularities.
However, this self-similar behavior is only observed for a specific range of vdW forces. In other parameter regimes, the profiles do not exhibit clear self-similarity as predicted by the theoretical solution. This limitation primarily arises from the influence of cylindrical curvature, whose effect on nonlinear evolution dynamics requires more rigorous theoretical treatment in future works.

\section{Conclusion}
\label{sec:conclusion}

In this study, the influence of vdW forces on the instability of liquid films in cylindrical tubes is investigated systematically, encompassing both the linear stage of small perturbations and the nonlinear evolution preceding film rupture or collapse. The strengths of vdW forces from the solid–liquid and liquid–liquid interactions are characterized by the dimensionless Hamaker constants $A_1$ and $A_2$, respectively.
A theoretical model is developed based on linear stability analysis of the Stokes equations, extending beyond classical lubrication theory. Results show that the lubrication model overestimates both the dominant wavenumber $k_{{max}}$ and the perturbation growth rate $\omega$, whereas the proposed Stokes model provides more accurate predictions of film instability. These theoretical findings, along with further dynamic analysis, are validated through direct numerical simulation of the NS equations.
When the initial film thickness $\epsilon$ exceeds a critical value $\epsilon_c$, the film may collapse and form liquid plugs. By modeling the film under quasi-static pressure equilibrium, we derive the maximum sustainable film volume and the theoretical critical thickness $\epsilon_c$. Simulation results confirm that the liquid-liquid vdW forces reduce $\epsilon_c$, enabling the collapse of thinner films.
During the linear stage of perturbation growth, both types of vdW forces increase $k_{{max}}$ and $\omega$, with ultra-thin films ($\epsilon < 0.1$) exhibiting particularly strong wavelength reduction. The predicted trend in dominant wavelength is further supported by experimental data from \cite{tomo2022}. Additionally, the times to rupture ($t_r$) and collapse ($t_c$) can be qualitatively predicted from linear growth rates.
In the nonlinear regime, vdW forces are shown to induce film rupture, motivating a comparative analysis of the morphology and scaling laws associated with rupture and collapse singularities. For rupture, increasing $A_1$ progressively suppresses the formation of satellite lobes by accelerating local thinning near the wall, with thicker films requiring stronger vdW forces to fully eliminate satellite lobes. Scaling analysis confirms that the minimum film thickness $1 - h_{{max}}$ follows a $\tau^{1/3}$ power law as $\tau = t_r - t \to 0$, consistent with dimensional analysis under viscosity- and vdW-dominated regimes. Similarly, during collapse governed by $A_2$, the minimum thickness $h_{{min}}$ adheres to the $\tau^{1/3}$ scaling with $\tau = t_c - t$.
Using the characteristic vdW length $a = \sqrt{\tilde{A} / 6\pi\gamma}$ to rescale the system, both rupture and collapse data collapse onto a unified $\tau^{1/3}$ scaling curve. 

In summary, a theoretical framework is established describing the evolution of liquid films under vdW forces across different stages in this work. We anticipate experimental validation of key predictions, particularly the relationships between the strength of vdW forces, dominant wavelength, and satellite lobe, through controlled destabilization of nanoscale films.
The proposed framework offers promising extensions, such as incorporating thermal fluctuation effects \citep{Fetzer2007,zhao2019} or substrate slip conditions \citep{liao2013,zhao2023}, both commonly encountered in nanoscale liquid film flows. We further expect that more detailed experimental observations of nanoscale liquid film dynamics will align with the theoretical predictions presented in this study.
\\

\begin{bmhead}[Acknowledgements.]
Useful discussions with Dr Z. Ding are gratefully acknowledged.
\end{bmhead}

\begin{bmhead}[Funding.]
This work has been supported by the National Natural Science Foundation of China (No. 12572310, 12202437, 12388101), the New Cornerstone Science Foundation through the XPLORER PRIZE, the Strategic Priority Research Program of the Chinese Academy of Sciences (XDB0910100), the Fundamental Research Funds for the Central Universities (WK2090000086), and the Startup Program of University of Science and Technology of China (KY2090000156).
\end{bmhead}

\begin{bmhead}[Declaration of interests.]
The authors report no conflict of interest.
\end{bmhead}

\begin{bmhead}[Author ORCIDs.]
\\Yixiao Mao https://orcid.org/0009-0001-7583-7740;
\\Chengxi Zhao https://orcid.org/0000-0002-3041-0882;
\\Yixin Zhang https://orcid.org/0000-0003-4632-3780;
\\Kai Mu https://orcid.org/0000-0002-4743-2332;
\\Ting Si https://orcid.org/0000-0001-9071-8646.
\end{bmhead}

\appendix

\section{Normal mode method of linear instability\label{appA}}
In this appendix, the instability analysis is performed for the Stokes equations using the normal mode method. Substituting the perturbed quantities (\ref{eq_smallap}) into the mass and momentum equations (\ref{eq_model4})--(\ref{eq_model6}) leads to
\begin{align}
\label{eq_stokes_linear3}
\iu k \hat{w} & + \frac{1}{r}  \frac{\du (\hat{u} r)}{\du r}  = 0 \,, \\
\label{eq_stokes_linear1}
\iu k \hat{p} & =  -k^2 \hat{w} + \frac{1}{r}\frac{\du}{\du r} \left( r \frac{\du \hat{w}}{\du r} \right) \,, \\
\label{eq_stokes_linear2}
\frac{\du \hat p}{\du r} & =  -k^2 \hat{u} + \frac{\du}{\du r}\left[\frac{1}{r} \frac{\du (\hat{u} r)}{\du r}\right] \,. 
\end{align}
With equations\,(\ref{eq_stokes_linear3})--(\ref{eq_stokes_linear2}), we eliminate $\hat{w}$ and $\hat{p}$ to derive a fourth-order ordinary differential equation for $\hat{u}$, as
\begin{equation}
\label{eq_stokes_linear4}
\frac{\du }{\du r}\left\lbrace\frac{1}{r} \frac{\du}{\du r} \left[r \frac{\du}{\du r} \left(\frac{1}{r} \frac{\du(\hat{u}r)}{\du r} \right) \right] \right\rbrace - 2k^2 \frac{\du}{\du r}\left[\frac{1}{r} \frac{\du(\hat{u}r)}{\du r} \right]+k^4\hat{u} = 0 \,.
\end{equation}
The general solution to (\ref{eq_stokes_linear4}) can be obtained in terms of Bessel functions:
\begin{equation}
\label{eq_stokes_solution1}
\hat{u} = C_1 r K_0(kr) + C_2 K_1(kr)+ C_3 r I_0(kr) + C_4 I_1(kr) \,,
\end{equation}
where $I_n$ and $K_n$ are modified Bessel functions of the first and second kind, respectively, with subscripts denoting their order. 
The constants $C_1\text{--}C_4$ are determined by boundary conditions, and $r\in[\alpha,1]$, where $\alpha=r_f/r_0$ is the dimensionless film position.
Substituting (\ref{eq_stokes_solution1}) into (\ref{eq_stokes_linear3}) and (\ref{eq_stokes_linear1}) gives
\begin{align}
\label{eq_stokes_solution2}
\hat{w}  = [C_1 \left(  krK_1-2K_0 \right)+ C_2& kK_0 - C_3\left(2 I_0 + kr I_1 \right) - C_4 kI_0] / (\iu k) \,, \\
\label{eq_stokes_solution3}
\hat{p} = 2& \left( C_1 K_0 + C_3 I_0 \right) \,.
\end{align}

In a similar approach, the perturbed quantities for $\hat{u}, \hat{w}, \hat{p}$, combined with $h(z,t) = \alpha + \hat{h}\eu^{\omega t+\iu kz}$, are substituted into perturbed boundary equations (\ref{eq_model7})--(\ref{eq_bc1}). 
Linearising the boundary conditions at $r=\alpha$ yields: 
\begin{align}
\label{eq_stokes_bc3}
 &\frac{\du \hat{w}}{\du r} + \iu k \hat{u}  = 0\,, \\
\label{eq_stokes_bc4}
 \hat{p} - 2 \frac{\du \hat{u}}{\du r}  
 = \hat{h}&\left(-k^2+ \frac{1}{\alpha^2}+\frac{3A_1}{ (1-\alpha)^4} + \frac{3A_2}{8 \alpha^4} \right), \\
\label{eq_stokes_bc5}
 &\qquad\omega \hat{h} = \hat{u} \,.
\end{align}
And at the tube wall $r=1$, the no-slip and no-penetration conditions linearise to
\begin{align}
\label{eq_stokes_bc1}
\hat{u} = 0\,, \,\hat{w} = 0 \,.
\end{align}
According to equation\,(\ref{eq_stokes_bc5}), $\hat{h}$ in (\ref{eq_stokes_bc4}) can be eliminated, reducing the system to four boundary conditions: equations (\ref{eq_stokes_bc3}), (\ref{eq_stokes_bc4}), and (\ref{eq_stokes_bc1}). 
Substituting the Bessel functions solutions (\ref{eq_stokes_solution1})--(\ref{eq_stokes_solution3}) into these perturbed equations  yields a homogeneous linear system for the coefficients $C_1\sim C_4$. For non-trivial solutions to exist, the determinant of the coefficients must vanish:

\begin{equation}
\begin{array}{|cccc|}
F_{11} & F_{12}  & F_{13} & F_{14} \\ 
2K_0(k)-kK_1(k) & -kK_0(k)  & 2 I_0(k)+k I_1(k) & kI_0(k) \\ 
 K_0(k) & K_1(k) &  I_0(k) & I_1(k) \\
k \alpha K_0(k \alpha) - K_1(k\alpha) & k K_1(k\alpha) & k \alpha I_0(k \alpha) + I_1(k \alpha) & k I_1(k\alpha) \\
\end{array} =0 \,, 
\end{equation}
where
\begin{align}
F_{11} & = [-k^2 + 1/\alpha^2+3A_1/(1-\alpha)^4 +3A_2/8\alpha^4 ] \alpha K_0(k \alpha) - 2 \omega k \alpha K_1(k \alpha)\,, \nonumber \\
F_{12} & =  [-k^2 + 1/\alpha^2+3A_1/(1-\alpha)^4 +3A_2/8\alpha^4 ] K_1(k \alpha) - 2 \omega \left[ k K_0(k \alpha) + K_1(k \alpha)/\alpha \right]\,, \nonumber \\
F_{13} & = [-k^2 + 1/\alpha^2+3A_1/(1-\alpha)^4 +3A_2/8\alpha^4 ]\alpha I_0(k\alpha) + 2 \omega k \alpha I_1(k\alpha)  \,, \nonumber \\
F_{14} & = [-k^2 + 1/\alpha^2+3A_1/(1-\alpha)^4 +3A_2/8\alpha^4 ] I_1(k \alpha)  + 2 \omega \left[ kI_0(k \alpha) - I_1(k \alpha)/\alpha \right]\,. \nonumber
\end{align}
Because $\omega$ only appears linearly in the first line of the determinant, the dispersion relation between $\omega$ and $k$ can be expressed explicitly as
\begin{align}
\omega &= -\frac{1}{2}\left( k^2- \frac{1}{\alpha^2} - \frac{3A_1}{ (1-\alpha)^4} - \frac{3A_2}{8 \alpha^4} \right) \nonumber\\
&\times \frac{K_0(k \alpha) \alpha \Delta_1 - K_1(k \alpha) \Delta_2 + I_0(k \alpha) \alpha \Delta_3 -I_1(k \alpha) \Delta_4}{k \alpha K_1(k \alpha) \Delta_1- \left[ kK_0(k \alpha)+K_1(k \alpha)/\alpha \right] \Delta_2 -k \alpha I_1(k \alpha) \Delta_3 + 
\left[ k I_0(k \alpha) -I_1(k \alpha)/\alpha \right] \Delta_4} \,,
\end{align} 
where
\begin{align}
\Delta_1& = 
\begin{array}{|ccc|}
 -kK_0(k)  & 2 I_0(k)+k I_1(k) & kI_0(k) \\ 
 K_1(k) &  I_0(k) & I_1(k) \\
 k K_1(k\alpha) & k \alpha I_0(k \alpha) + I_1(k \alpha) & k I_1(k\alpha) \\
\end{array}\,, \nonumber \\ \nonumber
\Delta_2 &= 
\begin{array}{|ccc|}
2K_0(k)-kK_1(k)   & 2 I_0(k)+k I_1(k) & kI_0(k) \\ 
 K_0(k) &  I_0(k) & I_1(k) \\
k \alpha K_0(k \alpha) - K_1(k\alpha) & k \alpha I_0(k \alpha) + I_1(k \alpha) & k I_1(k\alpha) \\
\end{array}\,, \\ \nonumber
\Delta_3 &= 
\begin{array}{|ccc|}
2K_0(k)-kK_1(k) & -kK_0(k)  & kI_0(k) \\ 
 K_0(k) & K_1(k) & I_1(k) \\
k \alpha K_0(k \alpha) - K_1(k\alpha) & k K_1(k\alpha) & k I_1(k\alpha) \\
\end{array}\,, \\ \nonumber
\Delta_4 &= 
\begin{array}{|ccc|}
2K_0(k)-kK_1(k) & -kK_0(k)  & 2 I_0(k)+k I_1(k) \\ 
 K_0(k) & K_1(k) &  I_0(k) \\
k \alpha K_0(k \alpha) - K_1(k\alpha) & k K_1(k\alpha) & k \alpha I_0(k \alpha) + I_1(k \alpha)  \\
\end{array}\,.  
\end{align}

\section{Lubrication model of linear instability\label{appB}}
In this appendix, the governing equations and corresponding dispersion relations for the lubrication model in a cylindrical tube are performed. To get the lubrication equations from the axisymmetric NS equations, we need to establish the leading-order terms by their asymptotic expansion of $\varepsilon$, where $\varepsilon \ll 1$. The parameters are rescaled as
\begin{equation}
    \tilde{r}=\varepsilon\tilde{\lambda}r,\,\tilde{z}=\tilde{\lambda}z,\,
    \tilde{u}=\frac{\varepsilon^4\gamma}{\mu}u,\,\tilde{w}=\frac{\varepsilon^3\gamma}{\mu} w,\,
    \tilde{t}=\frac{\tilde{\lambda}\mu}{\varepsilon^3 \gamma}t,\,\tilde{p}=\frac{\varepsilon\gamma}{\tilde{\lambda}}p.
\end{equation}
Here, $\tilde{\lambda}=r_0/\varepsilon$ represents the radial characteristic length. The Stokes equations (\ref{eq_model2})--(\ref{eq_model3}) yields
\begin{align}
\label{eq_lmodel1}
\frac{\partial w}{\partial z}&+ \frac{1}{r} \frac{\partial(ur)}{\partial r} =0\,, \\
\label{eq_lmodel2}
\frac{\partial p}{\partial z}= \varepsilon^2&\frac{\partial^2 w}{\partial z^2}+ \frac{1}{r}\frac{\partial}{\partial r}\biggl(r \frac{\partial w}{\partial r}\biggr)\, ,\\ 
\label{eq_lmodel3}
\frac{\partial p}{\partial r}=\varepsilon^4 & \frac{\partial^2 u}{\partial z^2}+ \varepsilon^2\frac{\partial}{\partial r}\biggl[\frac{1}{r} \frac{\partial (wr)}{\partial r}\biggr]\,.
\end{align}
At the liquid--gas interface $r=h$, the boundary condition (\ref{eq_model7}) and stress balance equations (\ref{eq_model8})--(\ref{eq_model9}) yield
\begin{align}
\label{eq_lmodel4}
\frac{\partial h}{\partial t}+ w \frac{\partial h}{\partial z} =u\,, \\
\label{eq_lmodel5}
p-\frac{2\varepsilon^2}{1+\varepsilon^2(\partial_z h)^2} \biggl[ \frac{\partial u}{\partial r}- \frac{\partial h}{\partial z}\biggl(\frac{\partial w}{\partial r}+\varepsilon^2&\frac{\partial u}{\partial z} \biggr)+ \varepsilon^2\biggl( \frac{\partial h}{\partial z}\biggr)^2\frac{\partial w}{\partial z}\biggr]\nonumber\\
=\frac{\partial_z^2 h}{[1+\varepsilon^2(\partial_z h)^2]^{3/2}}-\frac{1}{\varepsilon^2 h\sqrt{1+(\partial_z h)^2}} &+\frac{A_1}{(1-h)^3} -\frac{A_2}{(2h)^3}\,, \\
\label{eq_lmodel6}
2 \varepsilon^2\frac{\partial h}{\partial z}\biggl( \frac{\partial u}{\partial r}- \frac{\partial w}{\partial z}\biggr) +\biggl[1-\varepsilon^2\biggl( \frac{\partial h}{\partial z}\biggr)^2&\biggr]\biggl( \frac{\partial w}{\partial r}+\varepsilon^2 \frac{\partial u}{\partial z}\biggr)=0\,. 
\end{align}
At the liquid-solid interface $r=1$, the boundary condition remains
\begin{equation}
    \label{eq_lmodel7}
    u=0\,,\,w=0.
\end{equation}
We filter out the higher-order terms of $\varepsilon$ from the governing equations above, leading to
\begin{align}
\label{eq_le1}
    \frac{\partial w}{\partial z}+\frac{1}{r}&\frac{\partial (wr)}{\partial r}=0\,,\\
    \label{eq_le2}
    0=-\frac{\partial p}{\partial z}+&\frac{1}{r\partial r}(r\frac{\partial w}{\partial r})\,,\\
    \label{eq_le3}
    0=&\frac{\partial p}{\partial r}\,,\\
    \label{eq_le_bc1}
    \frac{\partial h}{\partial t}+w\frac{\partial h}{\partial z}&-u=0\,,\\
    \label{eq_le_bc2}
    p=\frac{\partial^2 h}{\partial z^2}-\frac{1}{h}+\frac{A_1}{(1-r)^3}-&\frac{A_2}{(2r)^3}\,,\,\frac{\partial w}{\partial r}=0\,,\,\rm{when}\,\it{r=h}\,,\\
    \label{eq_le_bc3}
    u=0\,,\,w=0&\,,\,\rm{when}\, \it r=\rm1.
\end{align}
Integrating (\ref{eq_le2}) from $r = h$ to $r = r$ with the boundary condition equation (\ref{eq_le_bc2}) gives
\begin{equation}
\label{eq_le_in1}
    \frac{1}{2}\frac{\partial p}{\partial z}(r^2-h^2)=r\frac{\partial w}{\partial r}\,.
\end{equation}
After integrating (\ref{eq_le_in1}) from $r = 1$ to $r = r$ with the slip boundary condition (\ref{eq_le_bc3}), we have
\begin{equation}
\label{eq_le_in2}
    \frac{1}{2}\frac{\partial p}{\partial z}\biggl[\frac{1}{2}(r^2-1)-h^2\ln r \biggr]=w\,.
\end{equation}
Using (\ref{eq_le1}), (\ref{eq_le_bc1}) and (\ref{eq_le_bc3}) with the Leibniz integral rule, one can obtain
\begin{equation}
    h\frac{\partial h}{\partial t}=\frac{\partial}{\partial z}\int^1_h ur\,\rm{d}\it r\,.
\end{equation}
Substituting (\ref{eq_le_in2}) into the integral results in
\begin{equation}
\label{eq_le_in3}
    \frac{\partial h}{\partial t}=\frac{1}{h}\frac{\partial}{\partial z}\biggl[ M(h)\frac{\partial p}{\partial z} \biggr]\,,
\end{equation}
where
\begin{equation}
    M(h)=(-1+4h^2-3h^4+4h^4\ln h)/16\,.
\end{equation}
To compare the Stokes model (\ref{eq_dispersion_stokes}) with the lubrication theory, (\ref{eq_le_in3}) is linearized using $h(z,t) = \alpha + \hat{h}\eu^{\omega t+\iu kz}$, providing the dispersion relation
\begin{equation}
\label{eq_le_d}
    \omega=k^2 \biggl( k^2-\frac{1}{\alpha^2}-\frac{3A_1}{(1-\alpha)^4}-\frac{3A_2}{8\alpha^4} \biggr)M\,,
\end{equation}
where $M=(-1+4\alpha^2-3\alpha^4+4\alpha^4\ln \alpha)/16$.

\section{Equilibrium shape and critical volume of collars\label{appC}}
In this appendix, we analyze the solution of the liquid--film interface equilibrium equation under different conditions, including the reason why $A_1 > 0$ is not applicable and the detailed effects of $A_2$. The interface equation is given by
\begin{equation}
    p=\frac{h''}{(1+h'^2)^{3/2}}-\frac{1}{h\sqrt{1+h'^2}} +\frac{A_1}{(1-h)^3} -\frac{A_2}{(2h)^3}=const,
    \label{eq:pconst_a}
\end{equation}
which is derived under quasi-static equilibrium assumptions, and constitutes a second-order ordinary differential equation for $h$ with respect to $z$. Considering that adjacent liquid collars are connected by an ultra-thin film, we impose the following boundary conditions: at the edge of a collar (set at $z=0$) we have $h = 1$ and $h' = 0$, while the curvature-related term $h''$ is treated as a free parameter.

When $A_1 > 0$, setting $h(0) = 1$ leads to a singularity ($p \to \infty$), which is physically inconsistent for a stable collar. To approximate our case, we instead set $h(0) = 1 - 10^{-\beta}$. 
The resulting profiles within a wavelength exhibit strong dependence on $\beta$, diverging significantly from the actual physical behavior under investigation.
Moreover, since vdW forces intensify sharply as the film thins, our simulations confirm that liquid film rupture inevitably occurs at the thinnest point when $A_1 > 0$, precluding any stable equilibrium state. Consequently, the theoretical framework above is unsuitable for analyzing the influence of $A_1$ on the collapse thickness, and we set $A_1 = 0$ in the corresponding analysis.

\begin{figure}
    \centering
    \includegraphics[width=1\linewidth]{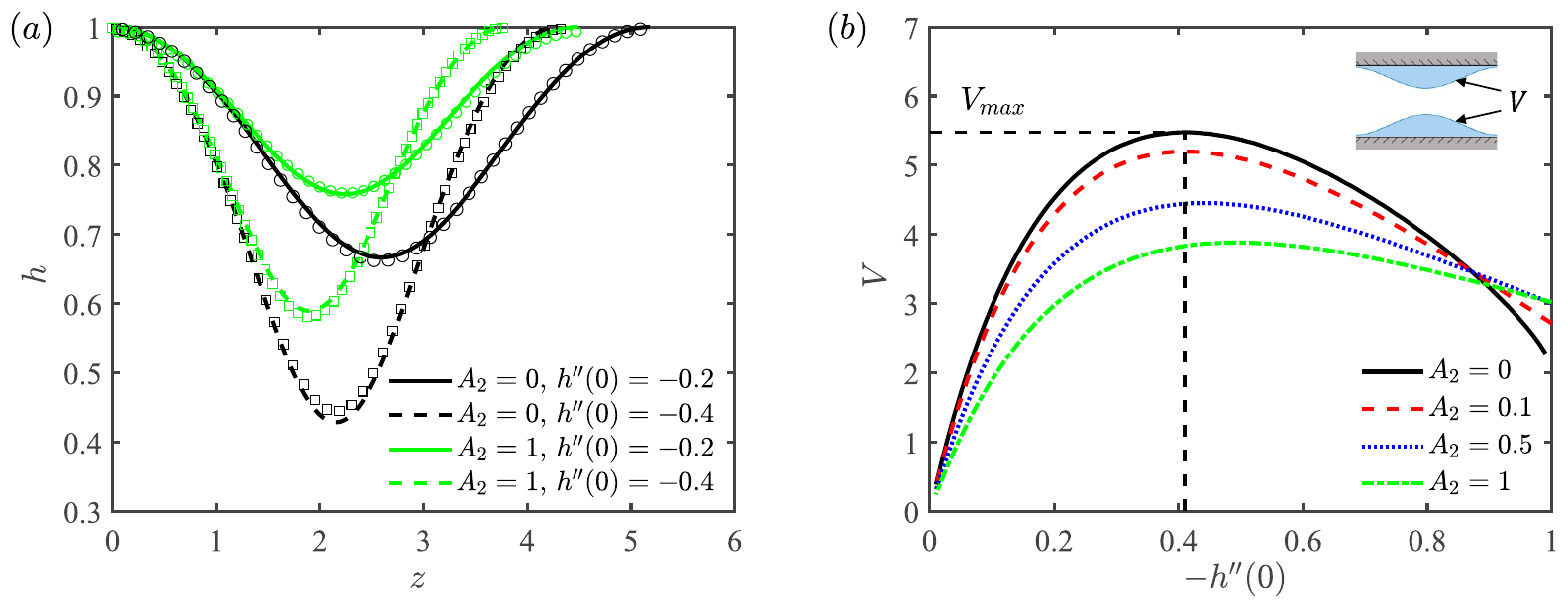}
    \caption{($a$) Equilibrium shape of a single collar in the case of different boundary conditions (as well as interface pressure). Lines represent theoretical results and symbols represent corresponding numerical profiles. ($b$) Liquid volume $V$ in a collar as a function of boundary condition $h''(0)$ for different $A_2$, where $V_{max}$ indicates the maximum volume.}
    \label{fig:collar shape}
\end{figure}

When considering the influence of $A_2$, different values of $h''$ yield distinct equilibrium interface profiles $h(z)$. As shown in figure \ref{fig:collar shape}($a$), when $A_2 = 0$, the interface with $h''(0) = -0.4$ displays a sharper contour than that with $h''(0) = -0.2$. In contrast, when $A_2 = 1$, the profiles become noticeably flatter. This flattening effect arises from vdW forces between liquid interfaces, which induce a radial contraction force dependent on the distance from the interface to the $z$-axis. To maintain pressure balance, the interface must adopt a more gradual curvature. 
Additionally, we compare the numerical profiles of stable collars obtained in the late stages of evolution under corresponding conditions. The results show good agreement with the theoretical predictions.

These morphological differences directly affect the film volume $V$, as depicted in figure \ref{fig:collar shape}($b$). For each vdW force condition, there exists a maximum sustainable volume $V_{{max}}$. Only films with volume per wavelength $V \leq V_{{max}}$ can maintain stable collar structures instead of collapsing. Furthermore, increasing $A_2$ reduces $V_{{max}}$, indicating that stronger intermolecular attractions enhance the film's tendency to collapse.

\clearpage

\bibliographystyle{jfm}
\bibliography{jfm}

@article{PRI,
   author = {Rayleigh, Lord},
   title = {On The Instability Of Jets},
   journal = {Proc. Lond. Math. Soc.},
   volume = {s1-10},
   number = {1},
   pages = {4-13},
   year = {1878}
}

@article{Dietze2015,
   author = {Dietze, G. F. and Ruyer-Quil, C.},
   title = {Films in narrow tubes},
   journal = {J. Fluid Mech.},
   volume = {762},
   pages = {68-109},
   year = {2015}
}

@article{Halpern2003,
   author = {Halpern, D. and Grotberg, J. B.},
   title = {Nonlinear saturation of the Rayleigh instability due to oscillatory flow in a liquid-lined tube},
   journal = {J. Fluid Mech.},
   volume = {492},
   pages = {251-270},
   year = {2003}
}

@article{Frenkel1987,
   author = {Frenkel, A. L. and Babchin, A. J. and Levich, B. G. and Shlang, T. and Sivashinsky, G. I.},
   title = {Annular flows can keep unstable films from breakup: Nonlinear saturation of capillary instability},
   journal = {J. Colloid Interface Sci.},
   volume = {115},
   number = {1},
   pages = {225-233},
   year = {1987}
}

@article{Rykner2024,
   author = {Rykner, M. and Saikali, E. and Bruneton, A. and Mathieu, B. and Nikolayev, V. S.},
   title = {Plateau–Rayleigh instability in a capillary: assessing the importance of inertia},
   journal = {J. Fluid Mech.},
   volume = {1001},
   pages = {A15},
   year = {2024}
}

@article{Lister2006,
   author = {Lister, J. R. and Rallison, J. M. and King, A. A. and Cummings, L. J. and Jensen, O. E.},
   title = {Capillary drainage of an annular film: the dynamics of collars and lobes},
   journal = {J. Fluid Mech.},
   volume = {552},
   pages = {311–343},
   year = {2006}
}

@article{Pahlavan2019,
   author = {Pahlavan, A. A. and Stone, H. A. and McKinley, G. H. and Juanes, R.},
   title = {Restoring universality to the pinch-off of a bubble},
   journal = {Proc. Natl. Acad. Sci. U.S.A.},
   volume = {116},
   number = {28},
   pages = {13780-13784},
   year = {2019}
}

@article{Olbricht1996,
   author = {Olbricht, W. L.},
   title = {Pore-Scale Prototypes of Multiphase Flow in Porous Media},
   journal = {Annu. Rev. Fluid Mech.},
   volume = {28},
   number = {1},
   pages = {187-213},
   year = {1996}
}

@article{Zoueshtiagh2014,
   author = {Zoueshtiagh, F. and Baudoin, M. and Guerrin, D.},
   title = {Capillary tube wetting induced by particles: towards armoured bubbles tailoring},
   journal = {Soft Matter},
   volume = {10},
   number = {47},
   pages = {9403-9412},
   year = {2014}
}

@article{Heil2008,
   author = {Heil, M. and Hazel, A. L. and Smith, J. A.},
   title = {The mechanics of airway closure},
   journal = {Respir. Physiol. Neurobiol.},
   volume = {163},
   number = {1-3},
   pages = {214-221},
   year = {2008}
}

@article{Goren1961,
   author = {Goren, S. L.},
   title = {The instability of an annular thread of fluid},
   journal = {J. Fluid Mech.},
   volume = {12},
   number = {2},
   pages = {309-319},
   year = {1961}
}

@article{Ruyer2008,
   author = {Ruyer-Quil, C. and Treveleyan, P. and Giorgiutti-Dauphin{\'e}, F. and Duprat, C. and Kalliadasis, S.},
   title = {Modelling film flows down a fibre},
   journal = {J. Fluid Mech.},
   volume = {603},
   pages = {431-462},
   year = {2008}
}

@article{Fetzer2007,
   author = {Fetzer, R. and Rauscher, M. and Seemann, R. and Jacobs, K. and Mecke, K.},
   title = {Thermal Noise Influences Fluid Flow in Thin Films during Spinodal Dewetting},
   journal = {Phys. Rev. Lett.},
   volume = {99},
   pages = {114503},
   year = {2007}
}

@article{Falk2010,
   author = {Falk, K. and Sedlmeier, F. and Joly, L. and Netz, R. R. and Bocquet, L.},
   title = {Molecular Origin of Fast Water Transport in Carbon Nanotube Membranes: Superlubricity versus Curvature Dependent Friction},
   journal = {Nano Lett.},
   volume = {10},
   number = {10},
   pages = {4067-4073},
   year = {2010}
}

@article{Day2015,
   author = {Day, R. W. and Mankin, M. N. and Gao, R. and No, Y.-S. and Kim, S.-K. and Bell, D. C. and Park, H.-G. and Lieber, C. M.},
   title = {Plateau–Rayleigh crystal growth of periodic shells on one-dimensional substrates},
   journal = {Nat. Nanotechnol.},
   volume = {10},
   number = {4},
   pages = {345-352},
   year = {2015}
}

@article{Everett1972,
   author = {Everett, D. H. and Haynes, J. M.},
   title = {Model studies of capillary condensation. I. Cylindrical pore model with zero contact angle},
   journal = {J. Colloid Interface Sci.},
   volume = {38},
   number = {1},
   pages = {125-137},
   year = {1972}
}

@article{Gauglitz1988,
   author = {Gauglitz, P. A. and Radke, C. J.},
   title = {An extended evolution equation for liquid film breakup in cylindrical capillaries},
   journal = {Chem. Eng. Sci.},
   volume = {43},
   number = {7},
   pages = {1457-1465},
   year = {1988}
}

@article{Hammond1983,
   author = {Hammond, P. S.},
   title = {Nonlinear adjustment of a thin annular film of viscous fluid surrounding a thread of another within a circular cylindrical pipe},
   journal = {J. Fluid Mech.},
   volume = {137},
   pages = {363-384},
   year = {1983}
}

@article{Lu2011,
   author = {Lu, Z. and Rath, C. and Zhang, G. and Kandlikar, S. G.},
   title = {Water management studies in PEM fuel cells, part IV: Effects of channel surface wettability, geometry and orientation on the two-phase flow in parallel gas channels},
   journal = {Int. J. Hydrog. Energy},
   volume = {36},
   number = {16},
   pages = {9864-9875},
   year = {2011}
}

@article{Liao2013,
   author = {Liao, Y.-C. and Li, Y.-C. and Wei, H.-H.},
   title = {Drastic Changes in Interfacial Hydrodynamics due to Wall Slippage: Slip-Intensified Film Thinning, Drop Spreading, and Capillary Instability},
   journal = {Phys. Rev. Lett.},
   volume = {111},
   number = {13},
   pages = {136001},
   year = {2013}
}

@article{Tomo2022,
   author = {Tomo, Y. and Nag, S. and Takamatsu, H.},
   title = {Observation of Interfacial Instability of an Ultrathin Water Film},
   journal = {Phys. Rev. Lett.},
   volume = {128},
   number = {14},
   pages = {144502},
   year = {2022}
}

@article{Bian2010,
   author = {Bian, S. and Tai, C. F. and Halpern, D. and Zheng, Y. and Grotberg, J. B.},
   title = {Experimental study of flow fields in an airway closure model},
   journal = {J. Fluid Mech.},
   volume = {647},
   pages = {391-402},
   year = {2010}
}

@article{zhao2023,
   author = {Zhao, C. and Zhang, Y. and Si, T.},
   title = {Slip-enhanced Rayleigh–Plateau instability of a liquid film on a fibre},
   journal = {J. Fluid Mech.},
   volume = {954},
   pages = {A46},
   year = {2023}
}

@book{plateau1873,
  title={Statique exp{\'e}rimentale et th{\'e}orique des liquides soumis aux seules forces mol{\'e}culaires},
  author={Plateau, J. A. F.},
  volume={2},
  year={1873},
  publisher={Gauthier-Villars}
}

@article{Beaty2022,
   author = {Beaty, E. and Lister, J. R.},
   title = {Nonuniversal Self-Similarity for Jump-to-Contact Dynamics between Viscous Drops under van der Waals Attraction},
   journal = {Phys. Rev. Lett.},
   volume = {129},
   pages = {064501},
   year = {2022}
}

@article{Moreno-Boza2020,
   author = {Moreno-Boza, D. and Martínez-Calvo, A. and Sevilla, A.},
   title = {Stokes theory of thin-film rupture},
   journal = {Phys. Rev. Fluids},
   volume = {5},
   number = {1},
   pages = {014002},
   year = {2020}
}

@article{Zhao2019,
   author = {Zhao, C. and Sprittles, J. E. and Lockerby, D. A.},
   title = {Revisiting the Rayleigh–Plateau instability for the nanoscale},
   journal = {J. Fluid Mech.},
   volume = {861},
   pages = {R3},
   year = {2019}
}

@article{Martinez2021,
   author = {Mart{\'i}nez-Calvo, A. and Moreno-Boza, D. and Sevilla, A.},
   title = {Non-linear dynamics and self-similarity in the rupture of ultra-thin viscoelastic liquid coatings},
   journal = {Soft Matter},
   volume = {17},
   number = {16},
   pages = {4363-4374},
   year = {2021}
}

@article{Perazzo2018,
   title = {Emulsions in porous media: From single droplet behavior to applications for oil recovery},
   journal = {Adv. Colloid Interface Sci.},
   volume = {256},
   pages = {305-325},
   year = {2018},
   author = {Perazzo, A. and Tomaiuolo, G. and Preziosi, V. and Guido, S.}
}

@article{deryagin1955,
  title={Definition of the concept of, and magnitude of the disjoining pressure and its role in the statics and kinetics of thin layers of liquids},
  author={Deryagin, B. V.},
  journal={Koll. Zhus.},
  volume={17},
  pages={191--197},
  year={1955}
}

@article{Craster2009,
   author = {Craster, R. V. and Matar, O. K.},
   title = {Dynamics and stability of thin liquid films},
   journal = {Rev. Mod. Phys.},
   volume = {81},
   number = {3},
   pages = {1131-1198},
   year = {2009}
}

@article{Zhang1999,
   author = {Zhang, W. W. and Lister, J. R.},
   title = {Similarity solutions for van der Waals rupture of a thin film on a solid substrate},
   journal = {Phys. Fluids},
   volume = {11},
   number = {9},
   pages = {2454-2462},
   year = {1999}
}

@article{Paulsen2012,
   author = {Paulsen, J. D. and Burton, J. C. and Nagel, S. R. and Appathurai, S. and Harris, M. T. and Basaran, O. A.},
   title = {The inexorable resistance of inertia determines the initial regime of drop coalescence},
   journal = {Proc. Natl. Acad. Sci. U.S.A.},
   volume = {109},
   number = {18},
   pages = {6857-6861},
   year = {2012}
}

@article{Martinez2020,
   author = {Mart{\'i}nez-Calvo, A. and Rivero-Rodr{\'i}guez, J. and Scheid, B. and Sevilla, A.},
   title = {Natural break-up and satellite formation regimes of surfactant-laden liquid threads},
   journal = {J. Fluid Mech.},
   volume = {883},
   pages = {A35},
   year = {2020}
}

@book{Gennes2004,
   author = {de Gennes, P. G. and Brochard-Wyart, F. and Qu{\'e}r{\'e}, D.},
   title = {Capillarity and Wetting Phenomena: Drops, Bubbles, Pearls, Waves},
   publisher = {Springer New York},
   year = {2004}
}

@book{Israelachvili2011,
   author = {Israelachvili, J. N.},
   title = {Intermolecular and surface forces},
   publisher = {Academic Press},
   year = {2011}
}

@article{Gennes1985,
   author = {de Gennes, P. G.},
   title = {Wetting: statics and dynamics},
   journal = {Rev. Mod. Phys.},
   volume = {57},
   number = {3},
   pages = {827-863},
   year = {1985}
}

@article{Tomotika1935,
   author = {Tomotika, S.},
   title = {On the instability of a cylindrical thread of a viscous liquid surrounded by another viscous fluid},
   journal = {Proc. R. Soc. Lond. A},
   volume = {150},
   number = {870},
   pages = {322-337},
   year = {1935}
}

@article{Curtis1982,
   title = {Dimensional analysis and the pi theorem},
   journal = {Linear Algebra Appl.},
   volume = {47},
   pages = {117-126},
   year = {1982},
   author = {Curtis, W. D. and Logan, J. D. and Parker, W. A.}
}

@article{zhang2021,
   author = {Zhang, Y. and Sprittles, J. E. and Lockerby, D. A.},
   title = {Thermal capillary wave growth and surface roughening of nanoscale liquid films},
   journal = {J. Fluid Mech.},
   volume = {915},
pages = {A135},
   year = {2021}
}

@article{zhao2022,
   author = {Zhao, C. and Liu, J. and Lockerby, D. A. and Sprittles, J. E.},
   title = {Fluctuation-driven dynamics in nanoscale thin-film flows: Physical insights from numerical investigations},
   journal = {Phys. Rev. Fluids},
   volume = {7},
   number = {2},
 pages = {024203},
   year = {2022}
}

@article{Quere1989,
   year = {1989},
   volume = {10},
   number = {4},
   pages = {335},
   author = {Qu{\'e}r{\'e}, D. and di Meglio, J.-M. and Brochard-Wyart, F.},
   title = {Making van der Waals Films on Fibers},
   journal = {Europhys. Lett.},
}

@article{ji2019,
   author = {Ji, H. and Falcon, C. and Sadeghpour, A. and Zeng, Z. and Ju, Y. S. and Bertozzi, A. L.},
   title = {Dynamics of thin liquid films on vertical cylindrical fibres},
   journal = {J. Fluid Mech.},
   volume = {865},
   pages = {303-327},
   year = {2019}
}

@article{chen2024,
   author = {Chen, H. and Wang, G. and An, T. and Yin, Z. and Fang, H.},
   title = {Modelling the first droplet emission from an electrified liquid meniscus hanging at the nozzle tip},
   journal = {J. Fluid Mech.},
   volume = {987},
pages = {A38},
   year = {2024}
}

@article{marco2022,
   author = {De Corato, M. and Tammaro, D. and Maffettone, P. L. and Fueyo, N.},
   title = {Retraction of thin films coated by insoluble surfactants},
   journal = {J. Fluid Mech.},
   volume = {942},
pages = {A55},
   year = {2022}
}

@article{ubal2014,
   author = {Ubal, S. and Grassia, P. and Campana, D. M. and Giavedoni, M. D. and Saita, F. A.},
   title = {The influence of inertia and contact angle on the instability of partially wetting liquid strips: A numerical analysis study},
   journal = {Phys. Fluids},
   volume = {26},
   number = {3},
pages = {032106},
   year = {2014}
}

@article{Burton2005,
   author = {Burton, J. C. and Waldrep, R. and Taborek, P.},
   title = {Scaling and Instabilities in Bubble Pinch-Off},
   journal = {Phys. Rev. Lett.},
   volume = {94},
   number = {18},
 pages = {184502},
   year = {2005}
}

@article{Ding2019,
title = {Thermocapillary effect on the dynamics of liquid films coating the interior surface of a tube},
journal = {Int. J. Heat Mass Trans.},
volume = {138},
pages = {524-533},
year = {2019},
author = {Ding, Z. and Liu, Z. and Liu, R. and Yang, C.}
}

@article{Zhang2024, 
title={Linear stability theory and molecular simulations of nanofilm dewetting with disjoining pressure, strong liquid–solid slip and thermal fluctuations}, 
volume={996}, 
journal={J. Fluid Mech.}, 
author={Zhang, Y.}, 
year={2024}, 
pages={A19}
}

@article{Ben‐Noah2023,
   author = {Ben‐Noah, I. and Friedman, S. P. and Berkowitz, B.},
   title = {Dynamics of Air Flow in Partially Water‐Saturated Porous Media},
   journal = {Rev. Geophys.},
   volume = {61},
   number = {2},
 pages = {e2022RG000798},
   year = {2023}
}

@article{Jacobsen2024,
   author = {Jacobsen, A. L. and Venturas, M. D. and Hacke, U. G. and Pratt, R. B.},
   title = {Sap flow through partially embolized xylem vessel networks},
   journal = {Plant Cell Environ.},
   volume = {47},
   number = {9},
   pages = {3375-3392},
   year = {2024}
}

@article{Erken2022,
   author = {Erken, O. and Romanò, F. and Grotberg, J. B. and Muradoglu, M.},
   title = {Capillary instability of a two-layer annular film: an airway closure model},
   journal = {J. Fluid Mech.},
pages = {A7},
   volume = {934},
   year = {2022}
}

@article{Beaty2023,
   author = {Beaty, E. and Lister, J. R.},
   title = {Inertial and viscous dynamics of jump-to-contact between fluid drops under van der Waals attraction},
   journal = {J. Fluid Mech.},
   volume = {957},
   pages = {A25},
   year = {2023}
}

@article{Haefner2015,
   author = {Haefner, S. and Benzaquen, M. and Baumchen, O. and Salez, T. and Peters, R. and McGraw, J. D. and Jacobs, K. and Raphael, E. and Dalnoki-Veress, K.},
   title = {Influence of slip on the Plateau-Rayleigh instability on a fibre},
   journal = {Nat. Commun.},
   volume = {6},
   pages = {7409},
   year = {2015}
}

@article{Tomo2018,
   author = {Tomo, Y. and Askounis, A. and Ikuta, T. and Takata, Y. and Sefiane, K. and Takahashi, K.},
   title = {Superstable Ultrathin Water Film Confined in a Hydrophilized Carbon Nanotube},
   journal = {Nano Lett.},
   volume = {18},
   number = {3},
   pages = {1869-1874},
   year = {2018}
}

@article{Chireux2018,
   author = {Chireux, V. and Protat, M. and Risso, F. and Ondarçuhu, T. and Tordjeman, P.},
   title = {Jump-to-contact instability: The nanoscale mechanism of droplet coalescence in air},
   journal = {Phys. Rev. Fluids},
   volume = {3},
   number = {10},
pages = {102001},
   year = {2018}
}

@article{Calvo2025,
   author = {Calvo-Rivera, A. and Moreno-Boza, D. and Sevilla, A.},
   title = {Axisymmetric dynamics and equilibria of annular liquid films coating a fiber},
   journal = {Phys. Fluids},
   volume = {37},
   number = {10},
 pages = {102120},
   year = {2025}
}

@article{Vrij1966,
   author = {Vrij, A.},
   title = {Possible mechanism for the spontaneous rupture of thin, free liquid films},
   journal = {Discuss. Faraday Soc.},
   volume = {42},
   pages = {23-33},
   year = {1966}
}



\end{document}